\documentclass[iop]{emulateapj}

\usepackage{color}
\usepackage{natbib}
\usepackage{amsmath}
\usepackage{gensymb}
\shorttitle{Mass-Radius Relation}
\shortauthors{Neil \& Rogers}

\begin{document}

\title{Host Star Dependence of Small Planet Mass-Radius Distributions}

\author{Andrew R. Neil and Leslie A. Rogers}
\affil{The Department of Astronomy and Astrophysics, University of Chicago,
    Chicago, IL 60637}

\begin{abstract}
The planet formation environment around M dwarf stars is different than around G dwarf stars. The longer hot protostellar phase, activity levels and lower protoplanetary disk mass of M dwarfs all may leave imprints on the composition distribution of planets.  We use hierarchical Bayesian modeling conditioned on the sample of transiting planets with radial velocity mass measurements to explore small planet mass-radius distributions that depend on host star mass. We find that the current mass-radius dataset is consistent with no host star mass dependence. These models are then applied to the \textit{Kepler} planet radius distribution to calculate the mass distribution of close-orbiting planets and how it varies with host star mass. We find that the average heavy-element mass per star at short orbits is higher for M dwarfs compared to FGK dwarfs, in agreement with previous studies. This work will facilitate comparisons between microlensing planet surveys and \textit{Kepler}, and will provide an analysis framework that can readily be updated as more M dwarf planets are discovered by ongoing and future surveys such as \textit{K2} and \textit{TESS}.
\end{abstract}


\section{Introduction}

The \textit{Kepler} survey \citep{BoruckiEt2011bApJ} has discovered thousands of transiting exoplanets, leading to the robust characterization of the radius distribution of small planets \citep{Burke2015ApJ, FultonEt2017ApJ}. Subsequent radial velocity follow-up of transiting planets from \textit{Kepler} and other surveys, as well as transit timing variations, have constrained planet masses and allowed exploration of the composition distribution of planets, often in the form of a mass-radius relationship. Constraining the composition distribution of planets provides insights into different formation pathways and the prevalence of Earth-like rocky planets.

A mass-radius relationship (hereafter M-R relation) is a key ingredient necessary for the comparison of different exoplanet populations. While \textit{Kepler}, and in the near future \textit{TESS} \citep{RickerEt2015JATIS}, are able to characterize the radii of planets, \textit{WFIRST}, a future microlensing survey, will only be able to characterize masses. If we wish to combine transit and microlensing surveys to constrain planet occurrence rates, an M-R relation is essential for translating masses into radii, and vice versa. For example, \citet{SuzukiEt2016ApJ} showed that the break in the mass-ratio function of microlensing planets from the \textit{MOA-II} survey occurs at a higher mass than the peak in the \textit{Kepler} mass distribution by converting \textit{Kepler} radii to masses with an M-R relation. However, \textit{WFIRST} is much more sensitive to planets around M dwarf hosts than \textit{Kepler}, due to the innate characteristics of the microlensing method. Given these disparate stellar distributions, if planet formation differs around M dwarfs compared to FGK dwarfs, this may manifest itself in the M-R relation. In this paper, we consider a mass-radius relationship that depends on host star mass.

\subsection{Planet formation around M dwarfs}

Important differences during the protostellar phase between M and FGK dwarfs could impact planet formation. Before reaching the main sequence, stars undergo a period of contraction whereby the luminosity declines several orders of magnitude. For low mass stars, this pre-main sequence phase lasts longer and the luminosity difference is more pronounced, leading to an initially distant snow line that migrates far inward while the star is contracting \citep{Kennedy&Kenyon2008ApJ}. This has important consequences for planet composition. \citet{RaymondEt2007ApJ} and \citet{Lissauer2007ApJ} found that planets formed in situ around low mass stars may be deficient in volatiles, as water-bearing planetesimals would not be scattered as often from beyond the snow line. However, if the surface density of water content in planetesimals is much higher than in our own solar system, volatile delivery to planets inside the snow line would increase \citep{CieslaEt2015ApJ} Alternatively, if the disk is long-lived, planets that form beyond the snow line would have more time to migrate inward, leading to volatile-rich planets at short orbits \citep{Alibert&Benz2017AAP}.

Another characteristic feature of M dwarfs is their intense activity at young stellar ages, during which flares can output bursts of XUV radiation and relativistic particles. For planets formed in situ, this high energy radiation could provide enough heating to evaporate the atmosphere and any water content on the surface \citep{ScaloEt2007AsBio}. Alternatively, for gaseous planets that form outside the snow line and migrate inward, XUV radiation from the star may be enough to fully evaporate their envelopes, resulting in ``habitable evaporated cores" \citep{LugerEt2015AsBio}. This process is more likely to occur for planets less massive than 2 $M_{\oplus}$, which could result in a population of volatile-rich rocky planets. If the mechanisms described can efficiently erode planet atmospheres, we expect close-in planets around M dwarfs to experience more mass loss and to be more dense than their counterparts around FGK stars.

\textit{Kepler} has shown that the radius distribution of planets around M dwarfs differs from that around FGK dwarfs, with higher occurrence rates of small planets around M dwarfs \citep{Dressing&Charbonneau2015ApJ, GaidosEt2016MNRAS}. Similarly, radial velocity surveys have shown differences in the the mass distribution, the most notable being a comparative lack of giant planets around M dwarfs \citep{BonfilsEt2013AAP}. Occurrence rate studies with \textit{Kepler} have also suggested a trend of increasing planetary heavy-element mass in short orbits for decreasing stellar mass, seemingly at odds with the protoplanetary disk mass scaling with stellar mass \citep{MuldersEt2015bApJ}. However, it is not yet clear as to how the composition distribution differs, or how this may be impacted by the stellar environment. If planets around M dwarfs have lower volatile content or experience more envelope mass loss, then we would expect planets to be more dense around low mass stars. However, this could be counteracted by a higher surface density of water content in the protoplanetary disk or longer disk lifetimes. While differences in the stellar environment exist for planets around different types of stars, there is currently no predictive model for how these differences may imprint themselves in the observed population of planets. Answering this question motivates characterizing more planets around M dwarf stars.

\subsection{M-R relations}

Empirically derived planet M-R relations have traditionally been cast as simple power laws. \citet{LissauerEt2011Nature} found $M/M_{\oplus} = (R/R_{\oplus})^{2.6}$ by fitting a power law to Earth and Saturn. Utilizing a sample of 22 planet pairs with TTV measured masses, \citet{Wu&Lithwick2013ApJ} found a linear relation with $M/M_{\oplus} = 3(R/R_{\oplus})$. \citet{WeissEt2013ApJ} introduced incident flux dependence into the M-R relation, using a sample of 35 planets with mass and radius measurements to find $(R/R_{\oplus}) = 1.78 (M/M_{\oplus})^{0.53} (F/ \text{erg}\ \text{s}^{-1}\ \text{cm}^{-2})^{-0.03}$ for $M < 150 M_{\oplus}$. \citet{Weiss&Marcy2014ApJL} used a sample of 42 \textit{Kepler} planets with RV-measured masses to fit a broken power law with a linear density relation $\rho_P = 2.43 + 3.39 (R/R_{\oplus})\  \text{g}\ \text{cm}^{-3}$ for $R/R_{\oplus} < 1.5$ and $M/M_{\oplus} = 2.69(R/R_{\oplus})^{0.93}$ for $1.5 < R/R_{\oplus} < 4.0$. 

More recently, \citet{WolfgangEt2016ApJ} used hierarchical Bayesian modeling to derive a more statistically robust, probabilistic M-R relation. They include intrinsic scatter in planet composition by modeling the M-R relation as a power-law with dispersion that is normally distributed. They find $M/M_{\oplus} = 2.7(R/R_{\oplus})^{1.3}$ for $R/R_{\oplus} < 4.0$ with an intrinsic scatter of $1.9 M/M_{\oplus}$, constrained to physically plausible densities, as the best-fit relation. The benefit of the Bayesian approach is that the uncertainties in the power law parameters and intrinsic scatter are fully quantified. \citet{ChenKipping2017ApJ} used a similar approach to fit broken probabilistic power laws to a much wider range of planet (and stellar) masses, including the transition points as additional parameters in the model. In this paper, we extend the approach by \citet{WolfgangEt2016ApJ} to include dependence on host star mass.

This paper is organized as follows. In section 2 we outline the steps taken to obtain our planet sample, detail our model and how we fit it to the data. The results of our model fitting are described in section 3, along with applications to planet heavy element mass and planet mass distributions. We discuss model selection, limitations and future extensions of this work in section 4, and conclude in section 5.

\section{Methods}

\subsection{Data}

In order to constrain the M-R relation, we need a sample of planets with well-characterized mass and radius measurements and uncertainties. For homogeneity, we choose to use the sample of transiting planets with radial velocity mass measurements. There is a comparable number of transiting planets with TTV mass measurements, but those planets have been shown to have systematically lower densities than planets with RV mass measurements \citep{Weiss&Marcy2014ApJL,JontofHutterEt2014ApJ}, likely due to different observational biases in the two techniques \citep{Steffen2016MNRAS, Mills2017ApJL}. We use only planets with radii below $8 R_{\oplus}$, as we are most interested in the M-R relation of small planets (rocky planets and mini-Neptunes / super-Earths). There is evidence for a transition in the M-R relation from Neptunian to Jovian planets at around ${\sim} 11 R_{\oplus}$ \citep{ChenKipping2017ApJ}, so adopting a cutoff of $8 R_{\oplus}$ is a conservative measure to ensure we are safely outside of the giant planet regime. Physically, as a planet becomes more massive, the core becomes more dense and degeneracy pressure plays an increasingly important role, causing the M-R relation to flatten at around a Jupiter radius \citep{Chabrier&Baraffe2007ApJ}.

The planet catalog for this work was downloaded from the NASA Exoplanet Archive\footnote{https://exoplanetarchive.ipac.caltech.edu/} on June 13, 2017. Our first cuts are made to exclude planets with $R > 8 R_{\oplus}$, circumbinary planets, and those without either transit or RV measurements. As our model incorporates host star uncertainties, we require that a planet has both a well-defined radial velocity semi-amplitude measurement ($K$) and a planet radius to host star radius ratio ($r = R_p/R_*$) measurement, with uncertainties. While transit depth may be seen as a more fundamental parameter derived from a transit light curve, stellar limb-darkening causes the transit depth to deviate from the simple $\delta = (\frac{R_p}{R_*})^2$ relation \citep{Mandel&Agol2002ApJL}, with the limb-darkening law and coefficients varying from star to star. To avoid having to treat limb-darkening for each individual star, we use the radius ratio instead of the transit depth as the primary transit parameter. For each planet, we verify the source paper for the RV measurement to ensure each value reported in the Exoplanet Archive is correct. We remove any planets without reported $K$ or $r$ values, or those with only upper limits. In the case of planets with transit depth measurements but no $r$ (HD 219134 b and c), we calculate $r$ from the transit depth assuming no limb darkening. For those planets without reported eccentricity or inclination measurements, we assume $e = 0$ and $i = 90^{\degree}$.

\begin{figure}
    \centering
    \includegraphics[width=\columnwidth]{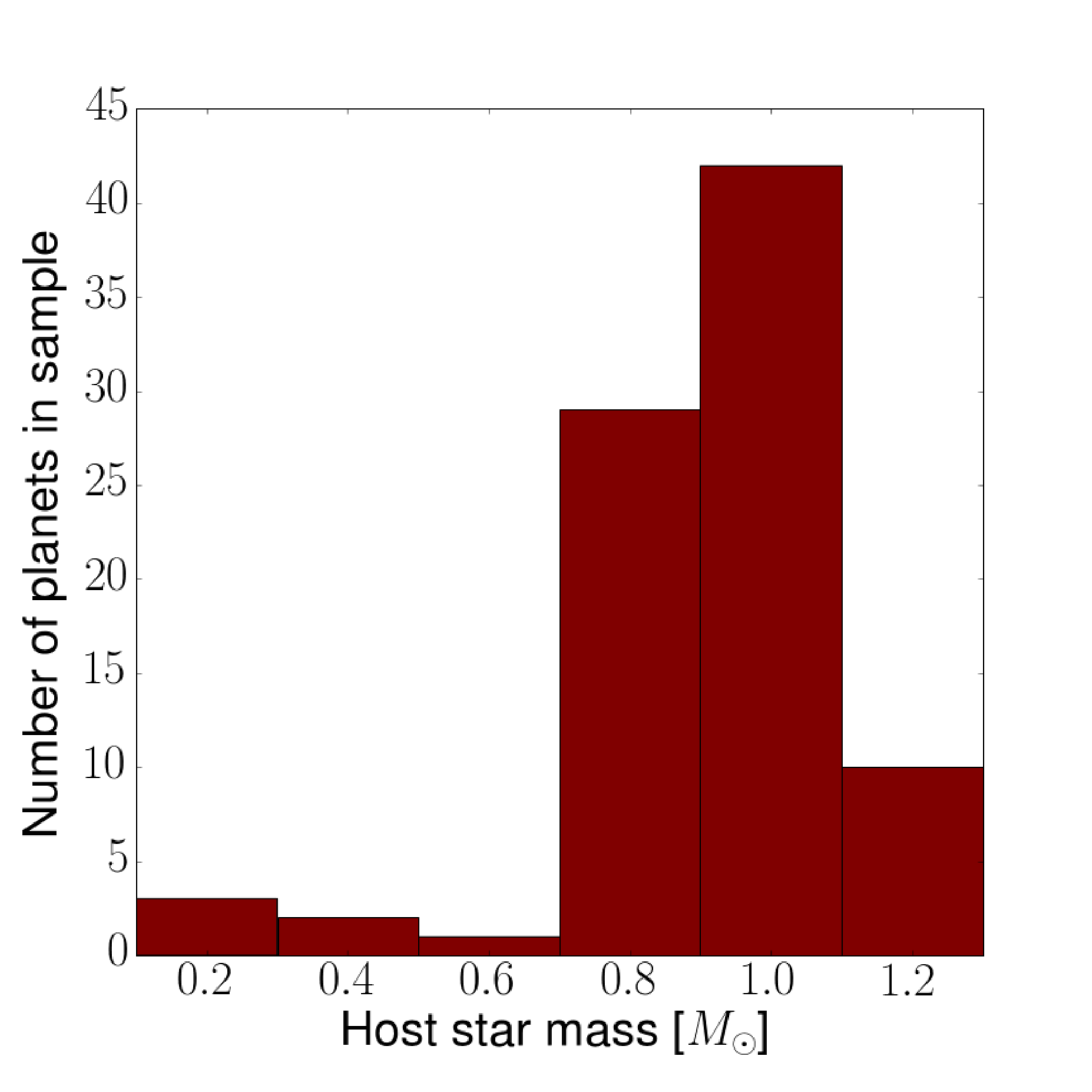}
    \caption{A histogram of planet host star mass in our sample. We use a sample of 87 transiting planets with radius $< 8 R_{\oplus}$ and RV-measured masses. Of those 87, only six orbit low-mass ($< 0.7 M_{\oplus}$) stars: GJ 436, GJ 1132, GJ 1214, GJ 3470, LHS 1140 and K2-18. The median host star mass of the sample is $0.95 M_{\odot}$.}
    \label{Planet host star mass histogram}
\end{figure}

The final planet catalog includes 88 planets. The distribution of host star mass for these planets is shown in Figure \ref{Planet host star mass histogram}. Given that we are most interested in how the M-R relation differs around stars of low mass, it is unfortunate that we only have six planets orbiting stars with $M < 0.7 M_{\odot}$; the majority of planets in the sample orbit G type stars. This is largely due to most radial velocity follow-up campaigns targeting GK dwarfs as solar analogues. However, there are many ongoing and future radial velocity observing programs that are specifically designed to characterize planets around M dwarfs (e.g.\textit{MAROON-X} \citep{SeifahrtEt2016}, \textit{SPIRou} \citep{MontouEt2015}, \textit{CARMENES} \citep{QuirrenbachEt2014}, \textit{ESPRESSO} \citep{GonzalezEt2017}) as a follow-up to transit observations (e.g. \textit{TESS}, \textit{MEarth} \citep{BertaEt2012}, \textit{ExTra} \citep{BonfilsEt2015SPIE}, \textit{SPECULOOS} \citep{BurdanovEt2017Arxiv}, \textit{APACHE} \citep{ChristilleEt2013}), so this number is expected to increase in the near future \citep{Kains}. Despite the current lack of planets with well-characterized radii and masses around M dwarfs, our analysis will provide a framework that can be revisited when the sample size increases drastically in the near future. A host star mass dependent M-R relation is also necessary to study trends in exoplanet populations with host star mass, and to allow comparison between \textit{Kepler} planets and planets discovered with the microlensing method.

\subsection{Model}

Since we are investigating the dependence of the M-R relation on host star mass, we choose to start with the same framework as \citet{WolfgangEt2016ApJ} in order to isolate this dependence. At a given radius, a planet's mass is drawn from a Gaussian distribution where the mean of the Gaussian distribution is a power-law:

\begin{equation} \label{eq:1}
\frac{M}{M_{\oplus}}  \sim \text{Normal}\left(\mu = C\left(\frac{R}{R_{\oplus}}\right)^\gamma, \sigma = \sigma_M\right)
\end{equation}

\noindent and the standard deviation parametrizes the intrinsic scatter. The fact that the mass is drawn from a distribution makes this a probabilistic, rather than deterministic, relation. 

In \citet{WolfgangEt2016ApJ}, $C$, $\gamma$ and $\sigma_M$ are the three parameters to be fit to the data. In our case, we expand each of these to include host star mass dependence:

\begin{equation} \label{eq:2}
\gamma = \gamma_0 + \ln{\left(\frac{M_*}{M_{\odot}}\right)} \gamma_s
\end{equation}

\begin{equation} \label{eq:3}
C = C_0 + \ln{\left(\frac{M_*}{M_{\odot}}\right)} C_s
\end{equation}

\begin{equation} \label{eq:4}
\sigma_M = \sqrt{\sigma_0^2 + \ln{\left(\frac{M_*}{M_{\odot}}\right)} \sigma_s}
\end{equation}

\noindent where we have introduced three new parameters that parameterize the host star dependence: $\gamma_s$, $C_s$, $\sigma_s$. In the case where these three parameters are all zero, then we obtain the original M-R relation in \citet{WolfgangEt2016ApJ}, which is independent of host star mass. In Equations (\ref{eq:2} - \ref{eq:4}), $M_*$ is the mass of the planet's host star, such that at a solar mass $M_{\odot}$, the host star mass dependent terms drop out, and the M-R relation is given solely by $\gamma_0$, $C_0$ and $\sigma_0$. Parameterizing the M-R relation in this manner allows both the mean planet mass and intrinsic scatter to smoothly vary as a function of host star mass. This specific choice of parametrization is chosen for simplicity, similar to the choice of parameterizing the M-R relation as a power-law. Our aim is to allow the M-R relation to change flexibly with host star mass, while still minimizing the number of parameters.

\citet{WolfgangEt2016ApJ} considered an additional model, where the intrinsic scatter in mass is allowed to vary as a function of planet radius. While this may be physically reasonable, they find that this parameter is consistent with zero and there is no strong evidence for requiring its inclusion. For this reason and for the sake of simplicity, we do not include any dependence of the intrinsic scatter on radius. However, following \citet{WolfgangEt2016ApJ}, we do include a maximum density constraint, based on the mass of a planet composed of a pure iron core. At a given planet radius, the maximum physically plausible planet mass is given by:

\begin{equation} \label{eq:5}
\log_{10}\left(M_{\text{pureFe}}\right) = \frac{-b + \sqrt{b^2 - 4a(c-R)}}{2a}
\end{equation}

\noindent where $a = 0.0975, b = 0.4938$ and $c = 0.7932$ \citep{FortneyEt2007bApJ}. Imposing this constraint has the effect of truncating the normal distribution of masses at a given radius. This is more constraining for small planet radii and severely limits the range of masses that these small planets can have. At around $1.5 R_\oplus$, this maximum mass limit no longer constrains the M-R relation significantly. In this way, although we have not included a parameter that allows the intrinsic scatter to vary with radius, the range of masses is being constrained at small radii by this maximum mass limit.

\begin{figure*}[ht]
    \centering
    \includegraphics[width=0.7\textwidth]{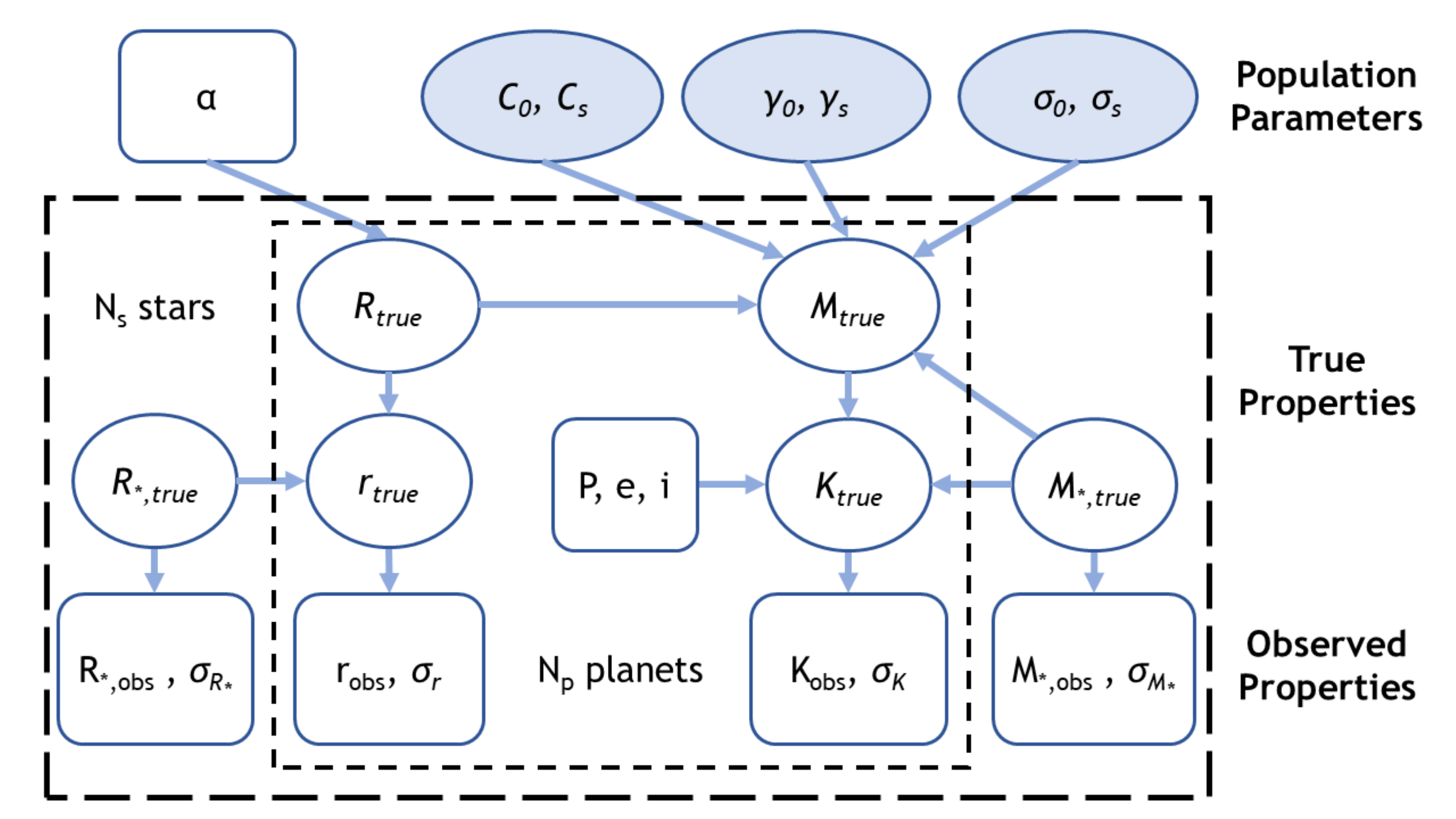}
    \caption{A graphical model of our host star mass dependent M-R relation as given by Equations (\ref{eq:1} - \ref{eq:8}). Rectangles represent input or assumed data; circles are parameters that are fitted for in our model. The shaded circles are the population level parameters of interest, whereas unshaded circles are inferred latent parameters.}
    \label{Planet mass radius graphical model}
\end{figure*}

A graphical representation of our model is shown in Figure \ref{Planet mass radius graphical model}. As input for our model, we take the following for each planet: observed radial velocity semi-amplitude $K_{\text{obs}}$ with uncertainty $\sigma_K$, observed planet radius to host star radius ratio $r_{\text{obs}}$ with uncertainty $\sigma_r$, period $P$, eccentricity $e$, and inclination $i$. We also take as input a list of observed host star masses $M_{*,\text{obs}}$ with uncertainty $\sigma_{M_*}$, as well as observed stellar radii $R_{*,\text{obs}}$ with uncertainty $\sigma_{R_*}$. We only include each host star once such that multiple planets around the same star will be drawing from the same host star properties during each step of the fitting. If asymmetric error bars were reported for a given measurement, we use the average of the two as the uncertainty. 

For each planet, we introduce the parameters $R_{\text{true}}$, $M_{\text{true}}$ which represent the true radius and mass of that planet. The true planet radii are assumed to follow some underlying distribution given by $\alpha$ (a uniform distribution for the purposes of this paper), while the true planet masses are given by our host star mass dependent M-R relation. The true planet masses and radii are both restricted to be above zero, and the true planet masses are restricted to be below the iron density constraint given above. We also introduce parameters corresponding to the true values of our observables: $K_{\text{true}}$, $r_{\text{true}}$, $M_{*,\text{true}}$, $R_{*,\text{true}}$. $K_{\text{true}}$ is calculated from the planet's period, inclination, eccentricity, true mass and host star mass while $r_{\text{true}}$ is calculated from the planet's true radius and host star radius, as shown below.

\begin{equation} \label{eq:6}
\begin{split}
K_{\text{true}} = \ & 0.6387 \left(\frac{P}{\text{day}}\right)^{-1/3} \left(\frac{M_{\text{true}}}{M_\oplus}\right) \frac{\sin{i}}{\sqrt{1-e^2}} \\
\times & \left(\left(\frac{M_{*,\text{true}}}{M_\odot}\right) + 3 \times{10^{-6}}\left(\frac{M_{\text{true}}}{M_\oplus}\right)\right)^{-2/3}\ [m/s] \\
r_{\text{true}} = \ & 0.009168 \left(\frac{R_{\text{true}}/R_\oplus}{R_{*,\text{true}}/R_\odot}
\right) \\
\end{split}
\end{equation}

Our full hierarchical model, including our list of priors for these parameters, is shown in Equations (\ref{eq:8} - \ref{eq:9}). U represents a uniform distribution, N represents a normal distribution, T represents truncation bounds, and $\sim$ represents ``is distributed as".

\begin{equation} \label{eq:8}
\begin{split}
K_{\text{obs}} & \sim \text{N}(K_{\text{true}}, \sigma_K) \\
r_{\text{obs}} & \sim \text{N}(r_{\text{true}}, \sigma_r) \\
M_{*,\text{obs}} & \sim \text{N}(M_{*,\text{true}}, \sigma_{M_*}) \\
R_{*,\text{obs}} & \sim \text{N}(R_{*,\text{true}}, \sigma_{R_*}) \\
M_{\text{true}} & \sim \text{N}(\mu, \sigma_M)\ T(0, M_{\text{true,Fe}})
\end{split}
\end{equation}

\begin{equation} \label{eq:9}
\begin{split}
C_0 & \sim \text{N}(5, 10)\ T(0,) \\
C_s & \sim \text{N}(0, 10) \\
\gamma_0 & \sim \text{N}(1, 1) \\
\gamma_s & \sim \text{N}(0, 1) \\
\sigma_0 & \sim \text{N}(3, 10)\ T(0,) \\
\sigma_s & \sim \text{N}(0, 20) \\
R_{\text{true}} & \sim \text{U}(0.2, 20) \\
M_{*,\text{true}} & \sim \text{U}(0.07, 2) \\
R_{*,\text{true}} & \sim \text{U}(0.05, 3) \\
\end{split}
\end{equation}

\subsection{Fitting the Model}

To fit our model to the data, we use the Python implementation of the Stan statistical software package \citep{CarpenterEt2017}\footnote{http://mc-stan.org}. Stan uses the No-U-Turn sampler (NUTS) MCMC algorithm, an efficient method of numerically evaluating hierarchical Bayesian models. For our host star mass dependent model, we ran 8 chains each with 30,000 iterations. The first 5000 iterations of each chain are thrown out as ``burn-in" to allow the chain to reach its equilibrium distribution, such that we are left with 200,000 posterior samples of each parameter. To assess the convergence and independence of each chain, we look at the output effective sample size (ESS) and Gelman-Rubin convergence diagnostic, $\hat{R}$. For each parameter, we ensure that $\hat{R} < 1.01$ and that the ESS is large. For our six parameters of interest, the ESS are all $>20,000$ while for our true planet and host star parameters, the ESS are all $>40,000$. Given these diagnostics, we assume the posteriors have converged to an acceptable degree.

\section{Results}

\subsection{Model fit}

\begin{table}
\centering
\begin{tabular}{c c c c} 
 \hline
 Parameter & 15.9\% & Median & 84.1\% \\
 \hline\hline
 $C_0$ & 2.45 & 2.97 & 3.52 \\ 
 $C_s$ & -1.77 & -0.26 & 0.67 \\
 $\gamma_0$ & 1.16 & 1.29 & 1.43 \\
 $\gamma_s$ & -0.09 & 0.22 & 0.61 \\
 $\sigma_0$ & 2.59 & 3.35 & 4.31 \\
 $\sigma_s$ & -12.06 & -4.18 & 1.39 \\
 \hline
\end{tabular}
\caption{Summary statistics of the six population level parameters from our model posteriors.}
\label{table:parameter posteriors}
\end{table}

\begin{figure*}[ht]
    \centering
    \includegraphics[width=0.8\textwidth]{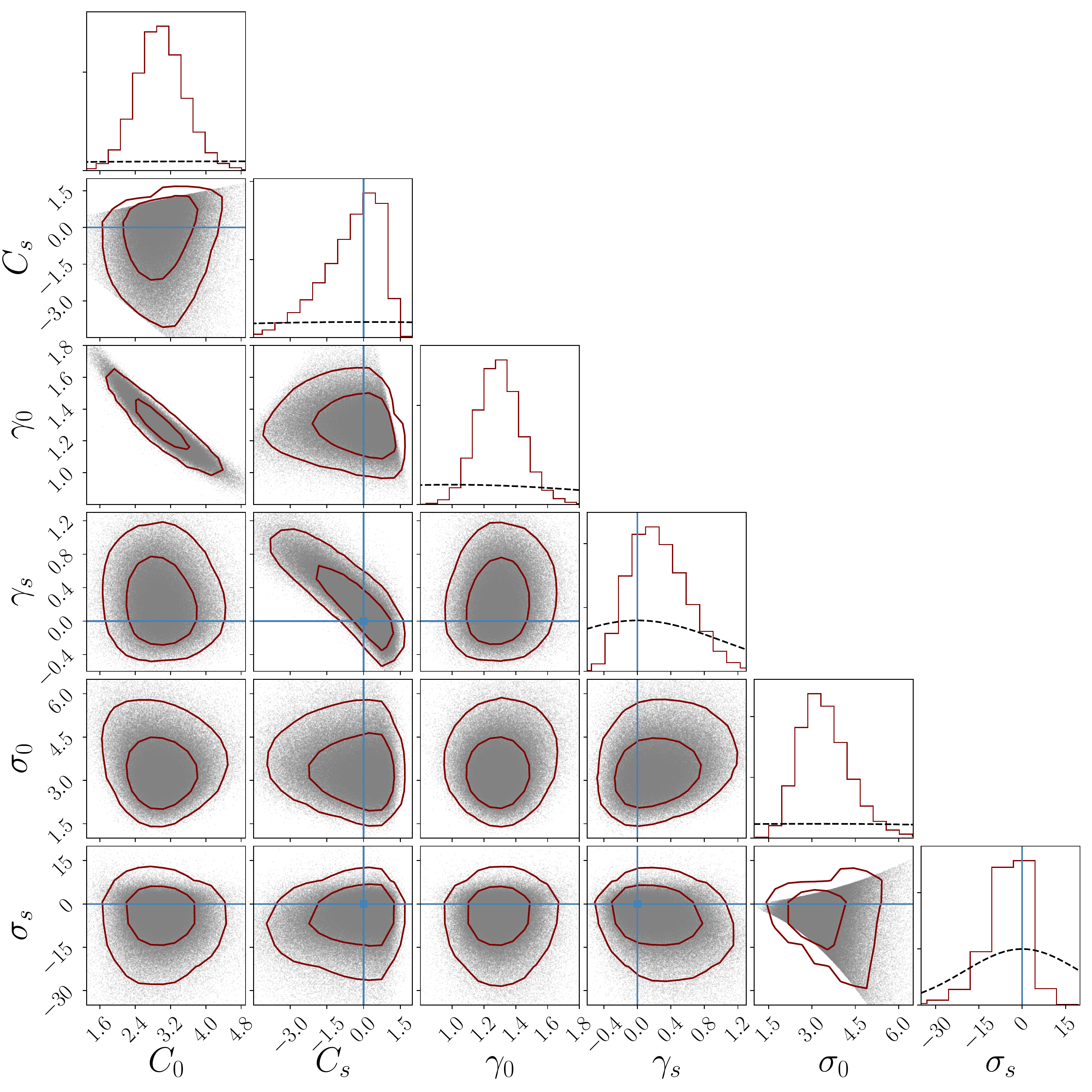}
    \caption{Posterior distributions from our six-parameter host star mass dependent M-R relation. In addition to power law slope, normalization, and intrinsic scatter, we introduce three parameters to allow the possibility of host star mass dependence. Contours show 68\% and 95\% confidence levels. All host star mass dependent parameters are consistent with zero (shown by the blue points), indicating that there is no robust evidence for host mass dependence in the current planet mass-radius dataset. The host star mass dependent intrinsic scatter, $\sigma_{s}$, with a median of -4.18 and a long negative tail, has the most support far from zero. Sharp ridges in the posterior distributions of $\sigma_{s}$ and $C_s$ arise from the constraint for $\sigma_M$ and $C$ to be positive. The black dotted lines in the 1D distributions indicate the priors, generally chosen to be weakly informative. Generated with corner.py \citep{corner}.}
    \label{Cornerplot}
\end{figure*}

Table \ref{table:parameter posteriors} summarizes the posteriors for our host star mass dependent M-R relation, and they are visualized in Figure \ref{Cornerplot}. We find that the three parameters encoding host star mass dependence $\left\{C_s, \gamma_s, \sigma_s\right\}$ are each consistent with zero. This can be seen in the set of 2D contour plots shown in Figure \ref{Cornerplot}, where the zero points for these three parameters (shown in blue) all lie within the $1\sigma$ contours. Further, the median values of $C_0$, $\gamma_0$ and $\sigma_0$ all closely match those found in \citet{WolfgangEt2016ApJ} and are much less sensitive to the priors than the three host star mass parameters, suggesting that the M-R relation does not have a strong dependence on host star mass.

Though the host star mass parameters are all consistent with zero, it is possible that the current mass-radius dataset is insufficient to robustly reveal any true host-star mass dependencies in the planet M-R distribution. Our analysis constrains the extent to which the host star mass parameters can deviate from zero; all the posterior distributions in Figure 3 are more strongly peaked than the assumed priors. The distribution of $\gamma_s$ peaks at ${\sim}$0.25, indicating slight preference for a steeper slope in the M-R relation towards higher host star masses, but is only discrepant from 0 at a $0.43\sigma$ level. The host star mass dependent scatter, $\sigma_s$, shows the most evidence for being discrepant from 0, with a median of -4.18 and 75\% of the posterior samples being below zero. This would suggest a larger scatter in the M-R relation toward lower host star masses. However, given that all three parameters do not rule out no host star mass dependence, we stress that this is only what the current dataset suggests with this model, and may not reflect underlying trends in the planet population. 

\begin{figure}
    \centering% \hspace{-0.5cm}
    \includegraphics[width=1\columnwidth]{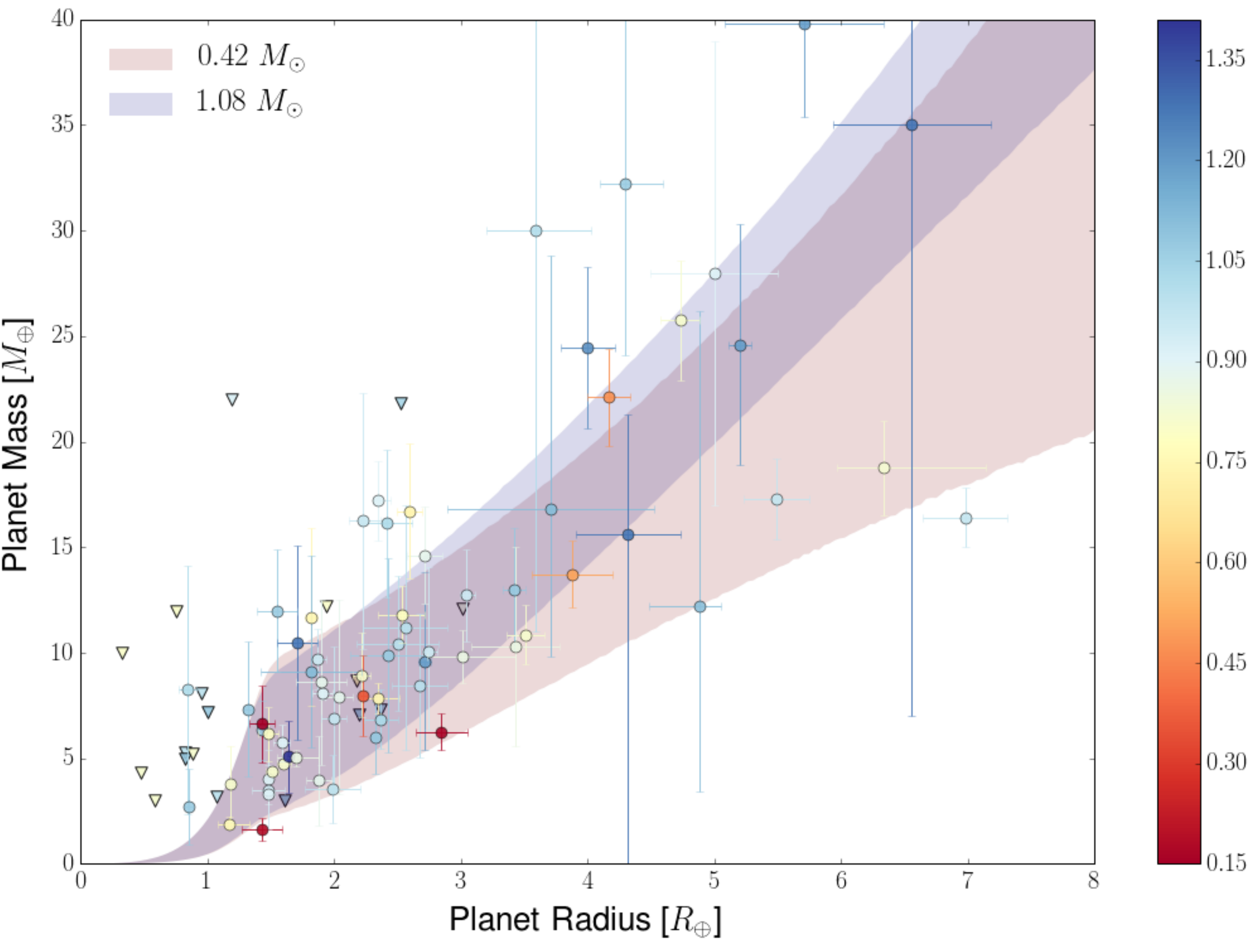}
    \caption{The host star mass dependent M-R relation for two characteristic stellar masses: 0.42 $M_{\odot}$, a typical M dwarf (red) and 1.2 $M_{\odot}$, a typical F dwarf (blue). The shaded region corresponds to the central 68\% of masses drawn at a given radius with the M-R relation parameters marginalized over their posterior distributions. The narrow feature below 1.5 $M_{\odot}$ is due to the pure iron maximum density restriction. The colored points represent the observed masses and radii of planets in our sample along with their reported uncertainties. The color of the points represents the host star mass as given by the colorbar, with redder points having lower host star mass and bluer points having higher host star mass. Triangles are upper limits in mass as reported by the original authors, although we use more complete information in our modeling.  Compared to the representative M-R relation for F stars, the M-R relation for M dwarfs has a shallower slope and higher intrinsic scatter. However, the posterior distributions for the host star mass dependent parameters are wide enough to be consistent with no host star mass dependence.}
    \label{Mass-radius relation F vs M}
\end{figure}

Figure \ref{Mass-radius relation F vs M} shows our host star mass dependent M-R relation marginalized over the posterior distribution for two different host star masses, representing M stars (in red) and F stars (in blue). As previously stated, the model prefers a shallower M-R relation with higher intrinsic scatter towards lower host star mass. This difference is most apparent at higher planet radii, with the spread of planet masses for rocky planets ($<1.5 R_\oplus$) being negligibly different between M and F host stars. Two of the most dominant limiting factors for this model are the small number of transiting planets with RV measured masses with $2.5 < R/R_{\oplus} < 8.0$, as well as the small number of those planets orbiting M dwarfs. This can be seen in Figure \ref{Mass-radius relation F vs M}, where the majority of planets fall between $1.5 R_\oplus$ and $2.5 R_\oplus$, and orbit solar-mass stars. These two limitations are largely due to the selection criteria of RV followup, where brighter (up to $T_{\text{eff}} \sim 6200$K) stars and lower radii planets are preferred, in the quest to find and characterize small, potentially rocky planets.

\subsection{Incident Flux Dependence}

As M dwarfs can have luminosities hundreds to thousands of times fainter than their GK counterparts, planets with the same period will have vastly different incident fluxes from their host star depending on the type of star. Overall, given similar period distributions, we expect planets around M dwarfs to have much lower incident fluxes. Incident flux has clear physical connections to planet composition. For example, there is a lack of ultra-short period planets with radius $<2 R_{\oplus}$, thought to arise from photoevaporation \citep{Sanchis-Ojeda2014ApJ} due to the extreme incident fluxes of these close-in planets. Evidence for photoevaporation can also be found in \textit{Kepler} radius distribution, which has been shown to have a gap at $\sim 1.8 R_{\oplus}$ \citep{FultonEt2017ApJ}.  Therefore, there is the possibility that any differences in the M-R relation due to host star mass can be attributed to different distributions of incident flux. We briefly explore incident flux dependence in the M-R relation and examine its effects on the host star mass dependence.

We model incident flux dependence into the M-R relation in an analogous fashion to the host star mass dependence in our primary model. We introduce three new parameters $\left\{C_f, \gamma_f, \sigma_f\right\}$ that modify the power-law slope, normalization and intrinsic scatter and are scaled by the natural log of the incident flux. Equations (2-4) then become:

\begin{equation} \label{eq:c incident flux}
C = C_0 + \ln{\left(\frac{M_*}{M_{\odot}}\right)} C_s + \ln{\left(\frac{S}{100 S_{\oplus}}\right)} C_f
\end{equation}

\begin{equation} \label{eq:gamma incident flux}
\gamma = \gamma_0 + \ln{\left(\frac{M_*}{M_{\odot}}\right)} \gamma_s + \ln{\left(\frac{S}{100 S_{\oplus}}\right)} \gamma_f
\end{equation}

\begin{equation} \label{eq:sigma incident flux}
\sigma_M = \sqrt{\sigma_0^2 + \ln{\left(\frac{M_*}{M_{\odot}}\right)} \sigma_s + \ln{\left(\frac{S}{100 S_{\oplus}}\right)} \sigma_f}
\end{equation}

\noindent where the incident flux $S$ for a given planet is calculated with the following equation:

\begin{equation}
\frac{S}{S_{\oplus}} = \left(\frac{R_*}{R_{\odot}}\right)^{2} \left(\frac{T_{\text{eff}}}{T_{\text{eff},\odot}}\right)^{4} \left(\frac{M_*}{M_{\odot}}\right)^{-2/3} \left(\frac{P}{P_{\oplus}}\right)^{-4/3}
\end{equation}

\noindent Our incident flux dependent model is then a nine parameter model, where each different combination of incident flux and host star mass gives a distinct M-R relation. 

\begin{figure}
    \centering
    \includegraphics[width=\columnwidth]{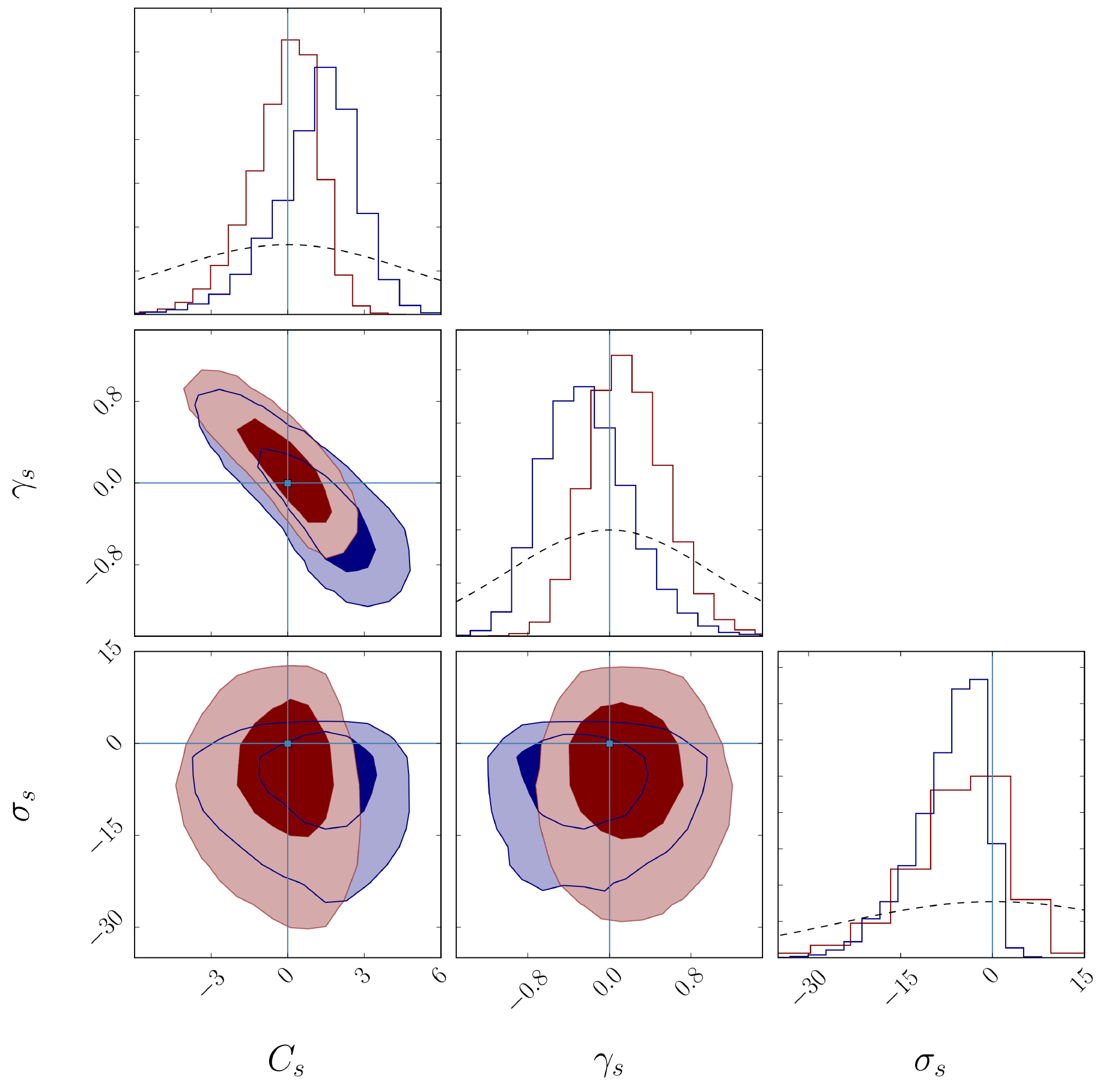}
    \caption{Posterior distributions from our nine parameter model that includes both incident flux and host star mass dependence in the M-R relation. Here we show only the host star mass dependent parameters, $\left\{C_s, \gamma_s, \sigma_s\right\}$. The posteriors from the six parameter model with only host star mass dependence are shown in red, while those from the nine parameter model with incident flux dependence are shown in blue, with the $1\sigma$ and $2\sigma$ contours shown. We find that the posteriors for $C_s$ and $\gamma_s$ widen when incident flux dependence is included, but the posterior distribution for $\sigma_s$ is slightly tighter. Additionally, the peak of the $\gamma_s$ distribution shifts from positive to negative, which corresponds to preferring a steeper slope in the M-R relation for low mass stars when incident flux dependence is included.}
    \label{incident flux posteriors}
\end{figure}

The posterior distributions for $\left\{C_s, \gamma_s, \sigma_s\right\}$ for our standard six parameter model compared to the nine parameter incident flux dependent model are shown in Figure \ref{incident flux posteriors}. We find that the posterior distributions for $C_s$ and $\gamma_s$ widen when incident flux dependence is introduced. We also note that while the original model prefers shallower slopes toward lower mass stars, with incident flux dependence this trend is reversed.  With the current dataset, it is difficult to disentangle the effects of incident flux and host star mass on the M-R relation. Both may affect planet composition, but currently there is no empirical proof for either affecting the M-R relation and no evidence to prefer one dependence over the other.

While this model does include both host star mass and incident flux dependence, this extra set of parameters is not justified by the limited dataset available. In section 4.1 we demonstrate by information criterion considerations that this model is not strongly preferred over the six parameter model. For this reason and for simplicity, we stick to the six parameter host star mass dependent M-R relation as the primary relation in the following sections. We have tested the conclusions of the following sections with the incident flux dependent model and have found the conclusions to be unchanged.

\subsection{Exploring Other Datasets}

\subsubsection{TTV Dataset}

As a consistency check, we repeat the modeling above using the TTV dataset instead of the RV dataset, as well as the combined RV + TTV datasets. We compiled the TTV dataset from the NASA Exoplanet Archive, with an identical cut on planet radius of $R < 8 R_{\oplus}$ as well as a cut on planet mass of $M < 25 M_{\oplus}$ to remove planets with physically unlikely masses. For TTV planets, we replace the radial velocity semi-amplitude in our model with the planet to host star mass ratio, $M / M_*$. We derive these values from the reported masses and host star masses, which is valid because host star uncertainties do not factor into TTV modeling \citep{HolmanEt2010Science}. For the combined RV and TTV dataset, if a planet has both RV and TTV mass measurements, we assume the RV measurement. The TTV dataset contains 63 planets compared to 88 RV planets, but has the advantage of more planets orbiting low-mass stars (13 planets with host star masses below $0.7 M_{\odot}$, compared to 6 planets with RV), although they come from fewer distinct systems. Due to overlap between the two datasets, the combined RV + TTV dataset has 143 planets.

\begin{figure}
    \centering
    \includegraphics[width=\columnwidth]{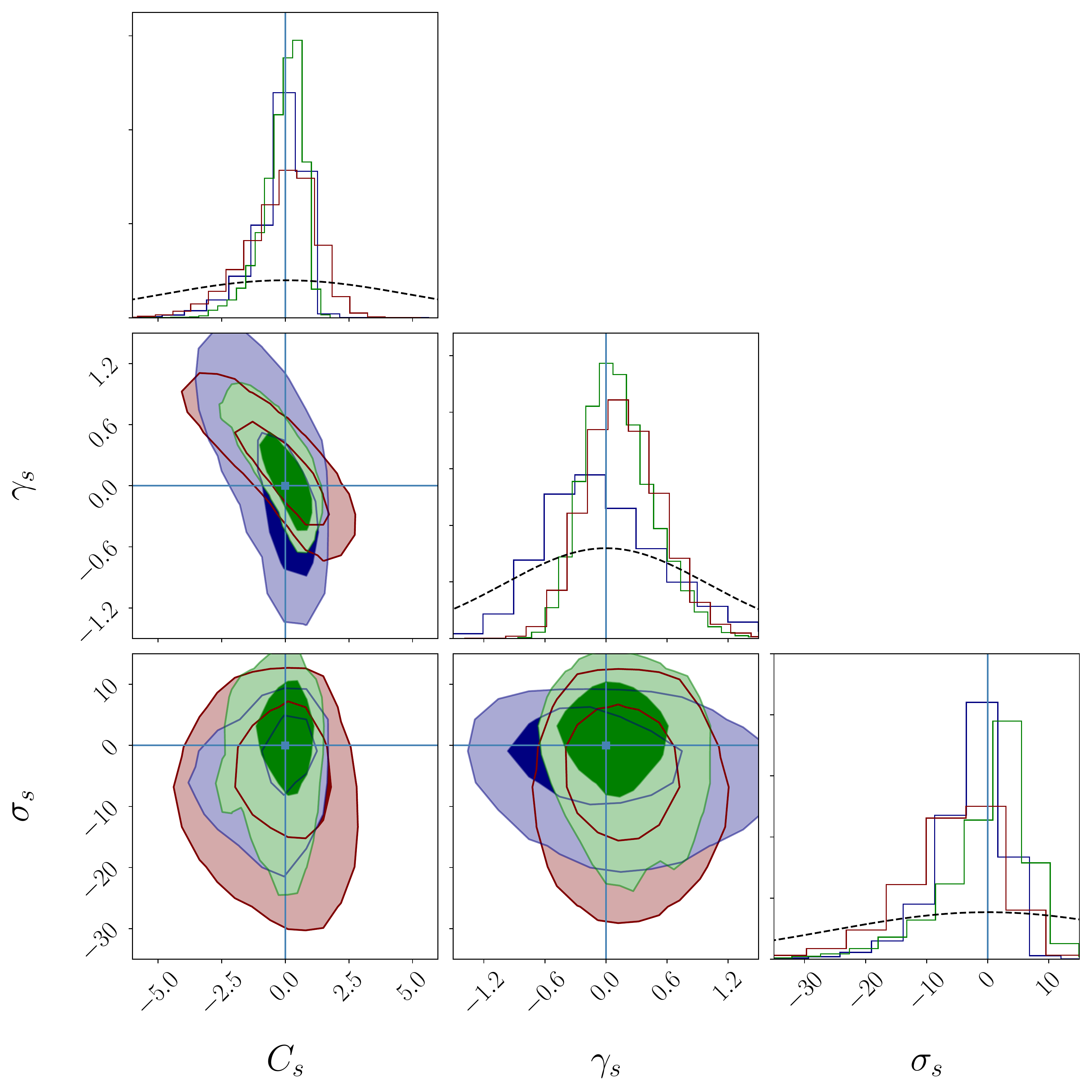}
    \caption{Posterior distributions from our six parameter model fit with three different datasets: RV (red), TTV (blue), and RV + TTV (green). As in Figure \ref{incident flux posteriors}, we show only the host star mass dependent parameters, $\left\{C_s, \gamma_s, \sigma_s\right\}$. We find that the posterior distributions agree between the three datasets, with the TTV data also showing no evidence for host star mass dependence.}
    \label{ttv posteriors}
\end{figure}

The posterior distributions of the three host star mass dependent parameters for our standard model fitted to the RV, TTV and RV + TTV datasets are shown in Figure \ref{ttv posteriors}. With the TTV dataset, we find that there is no evidence for host star mass dependence, as the posteriors for the three host star mass dependent parameters are consistent with zero. The median value of $\gamma_s$ using the TTV dataset is $-0.08$, which translates to a slightly shallower power law slope towards higher host star masses, a trend opposite to what we found with the RV dataset. Rather than revealing anything insightful about the underlying population, this is likely a result of our limited sample and further evidence for no host star mass dependence evident in the current sample of planets. The median value of $\sigma_s$ is $-1.88$ using the TTV dataset, in the same direction as the RV dataset. With the RV+TTV dataset, the posteriors tighten, but are still centered at 0.

\subsubsection{Potential TESS Dataset}

The lack of evidence for host star mass dependence in the M-R relation from fitting our model to the RV, TTV and RV + TTV datasets begs the question of how much data would be needed to demonstrate a dependence if such a dependence were to exist. The biggest hope for the near future is the \textit{TESS} survey and subsequent radial velocity follow-up, given the increased number of M dwarf planets \textit{TESS} expects to find compared to \textit{Kepler}. Using a simulated catalog of \textit{TESS} planets from \citet{Sullivan2015ApJ}, we construct a hypothetical future dataset of transiting planets with RV measured masses. We assume that each planet orbiting an M dwarf with a V magnitude above some cut will have a radial velocity mass measurement. We use a cut of $V_{\text{mag}} < 14$ for potentially rocky planets ($\frac{R_p}{R_\oplus} < 2$) and a cut of $V_{\text{mag}} < 13$ for those with gaseous envelopes ($2 < \frac{R_p}{R_\oplus} < 8$) to simulate the preference of many radial velocity follow-up programs to characterize rocky planets. We perform similar cuts on the FGK sample, using a cut of $K_{s,\text{mag}} < 10$ for rocky planets and $K_{s,\text{mag}} < 7$ for gaseous planets. We then generate radial velocity semi-amplitudes for each planet using two different models: the first using our six-parameter model with $\left\{C_0 = 3.0, C_s = 0.5, \gamma_0 = 1.3, \gamma_s = -0.5, \sigma_0 = 2.0, \sigma_s = -3.0\right\}$, and the second using a three-parameter model with no host star mass dependence $\left\{C = 3.0, \gamma = 1.3, \sigma = 2.0\right\}$. We make a further cut on the sample to ensure each planet has $K_\text{expected} > 1 \text{m/s}$ (taken from \citet{Sullivan2015ApJ}), in line with the sensitivity of radial velocity spectrographs such as \textit{MAROON-X} \citep{SeifahrtEt2016}. Finally, we generate normally distributed fractional errors for $K, r, R_*, M_*$ using values similar to those found in the current sample of small planets with mass and radius measurements. These mean fractional errors are 0.2, 0.015, 0.03, and 0.1, respectively. The true values of these parameters are perturbed with these errors to simulate our observables.

We fit both datasets to each of the two models used to generate them: the six-parameter model with host star mass dependence and the three-parameter model without. We do this for 10 realizations of each dataset. We then calculate the difference in WAIC (further discussed in section 4.1) for each dataset between the two models, as a means to test whether we can successfully distinguish which is the correct model. We find that this information criterion correctly prefers the model with no host star mass dependence for the dataset generated without host star mass dependence in 9 of 10 realizations. Further, compared to the smaller, current RV dataset, the $1\sigma$ uncertainties for the three host star mass dependent parameters decrease by a factor of $\sim 3$. For the dataset generated with host star mass dependence, we similarly find that 10 out of 10 realizations correctly preferred the model with host star mass dependence by at least $2\sigma$. We also note that 7 out of 10 realizations excluded 0 for $C_s$ at a $1\sigma$ level, with 10 out of 10 and 4 out of 10 realizations excluding zero for $\gamma_s$ and $\sigma_s$, respectively.

These results suggest that, while variance is still a factor, for this set of model parameters we would likely be able to find evidence for host star mass dependence with a future \textit{TESS} dataset. However, these results are contingent upon a few factors. While we generated 10 realizations of the radial velocity measurements, each realization used the same set of radii taken from the simulated catalog of \citet{Sullivan2015ApJ}. The true variance between datasets should be higher, if radii were generated alongside masses. We used a simplistic treatment of both the errors on the masses and radii, as well as the radial velocity follow-up strategy. The real catalog may look much different from the ones studied here. Further, we only tested one set of model parameters, and only one model parametrization with host star mass dependence: there are many plausible combinations of these parameters that may describe the underlying M-R relation. Despite these limitations, we present this as one method of determining whether or not \textit{TESS} may help us distinguish between M-R relations with and without host star mass dependence.

\subsection{Planet Mass Distributions}

The \textit{Kepler} survey has provided a wealth of potential planets, with over 4500 planet candidates discovered around a wide variety of host stars. This large sample of planet candidates has enabled many exoplanet population studies, including how exoplanet systems may differ around various types of stars. A pair of papers by \citet{MuldersEt2015aApJ,MuldersEt2015bApJ} (hereafter MPA15a,b) has provided evidence for increasing heavy-element mass in planetary systems in short orbits (period between 2-150 days) around lower-mass stars. They find that the average heavy-element mass in short orbits scales roughly inversely with host star mass, from 3.6 $M_{\oplus}$ around F stars to 7.3 $M_{\oplus}$ around M stars. This trend is at apparent odds with the observed trend of protoplanetary disk masses increasing with host star mass, derived from millimeter-wave observations \citep{AndrewsEt2013ApJ}.

As with any analysis of the \textit{Kepler} sample involving planet masses, the results in MPA15b are heavily reliant on the assumed M-R relation. They derived their results using both the deterministic M-R relation in \citet{Weiss&Marcy2014ApJL} as well as the probabilistic M-R relation in \citet{WolfgangEt2016ApJ}, and found the trend to be robust to the assumed M-R relation. However, they use the best-fit deterministic relation from \citet{WolfgangEt2016ApJ} rather than using the truly probabilistic formulation. More specifically, they did not take into account the intrinsic scatter of the M-R relation or the maximum density constraint derived from a pure iron core. This would have the effect of overestimating planet masses at small radii, which could skew the observed trend given that smaller planets have higher occurrence rates around lower mass stars. Additionally, uncertainties in the adopted M-R relation parameters are not taken into account, which would cause the errors reported to be underestimated. In this section, we seek to test the robustness of this trend, and to see to what extent it depends on the assumed planet M-R relation. We will use the posteriors from our host star mass dependent M-R relation for this purpose. We largely follow the methodology of MPA15b in deriving planet occurrence rates, which we briefly describe below.
 
For the purposes of this calculation, we use the Q1-16 KOI \citep{MullallyEt2015ApJS} catalog along with the corresponding list of target stars observed in these quarters. In order to isolate the effect of using our own M-R relation posteriors, we ensure that our KOI catalog completely matches that used in MPA15b. Giant stars are removed from the stellar catalog given their position in $\log{g}$-$T_{\mathrm{eff}}$ space according to the prescription in \citet{CiardiEt2011AJ}. Stellar noise during a transit is derived from the Combined Differential Photometric Precision (CDPP) metric \citep{ChristiansenEt2012PASP} at 3, 6 and 12 hour timescales for each observing quarter, which were downloaded from the MAST archive\footnote{http://archive.stsci.edu/kepler/data\_search/search.php}. For each star, we take the median of the CDPP of all quarters at each timescale and fit a power law. The CDPP for a given star and a given transit duration is then calculated using the star's CDPP power law fit.

The occurrence rate $f_{\text{occ}}$ of a given planet with radius and orbital period ${R_p, P}$ in a bin of stellar effective temperature $T_{\text{eff}}$ is defined as the inverse of the number of stars in that bin for which the planet would be detectable $N_*$, multiplied by a factor $f_{\text{geo}}$ to account for the geometric probability for the planet to transit:

\begin{equation}
f_{\text{occ}}(\left\{T_{\text{eff}}\right\}, R_p, P) = \frac{1}{f_{\text{geo}}N_*(\left\{T_{\text{eff}}\right\}, R_p, P)}
\end{equation}

\noindent We use $T_{\text{eff}}$ bins that correspond to M, K, G and F stars based on the recommendation of the Exoplanet Study Analysis Group (SAG) 13\footnote{https://exoplanets.nasa.gov/exep/exopag/sag/}, with upper bin edges of $3900$, $5300$, $6000$, and $7300$ K. For each star, the S/N of the planet is calculated given the transit depth, the number of transits and the stellar noise that the system would have with the planet orbiting that star. The detection efficiency given that S/N is then calculated based on the number of transits and a cumulative gamma function that is empirically derived from planet transit injection and recovery tests \citep{ChristiansenEt2015ApJ}. For detailed calculations, see equations (2-9) in MPA15a, as well as MPA15b.

In order to calculate planetary heavy-element mass as a function of host star mass, we first sample from our host star mass dependent M-R relation to obtain a set of the six parameters $\left\{C_0, C_s, \gamma_0, \gamma_s, \sigma_0, \sigma_s\right\}$. At the same time, we resample the KOI population by bootstrapping (drawing $N_p$ planets from the sample with replacement). Then, for each planet we sample $N_m$ masses using the set of host star mass dependent M-R relation parameters, as well as sampling the planet's radius and host star mass from normal distributions. To retrieve the heavy-element mass instead of the total mass, we cap the mean of the probabilistic M-R relation at $22 M_{\oplus}$ in order to replicate the measured median core masses of giant planets \citep{Miller&Fortney2011ApJ}. Each mass is sampled from a truncated normal distribution where the mean and standard deviation are given by our M-R relation, and the truncation is due to the maximum density of a planet with a pure iron core. For each sample of a planet's mass, we calculate the planet's contribution to the average heavy-element mass by multiplying its mass by the occurrence rate of the planet in its $T_{\text{eff}}$ bin. We repeat the outer level sampling of the M-R relation parameters and bootstrap resampling of the population $N_b$ times. 

\begin{figure}
    \centering
    \includegraphics[width=\columnwidth]{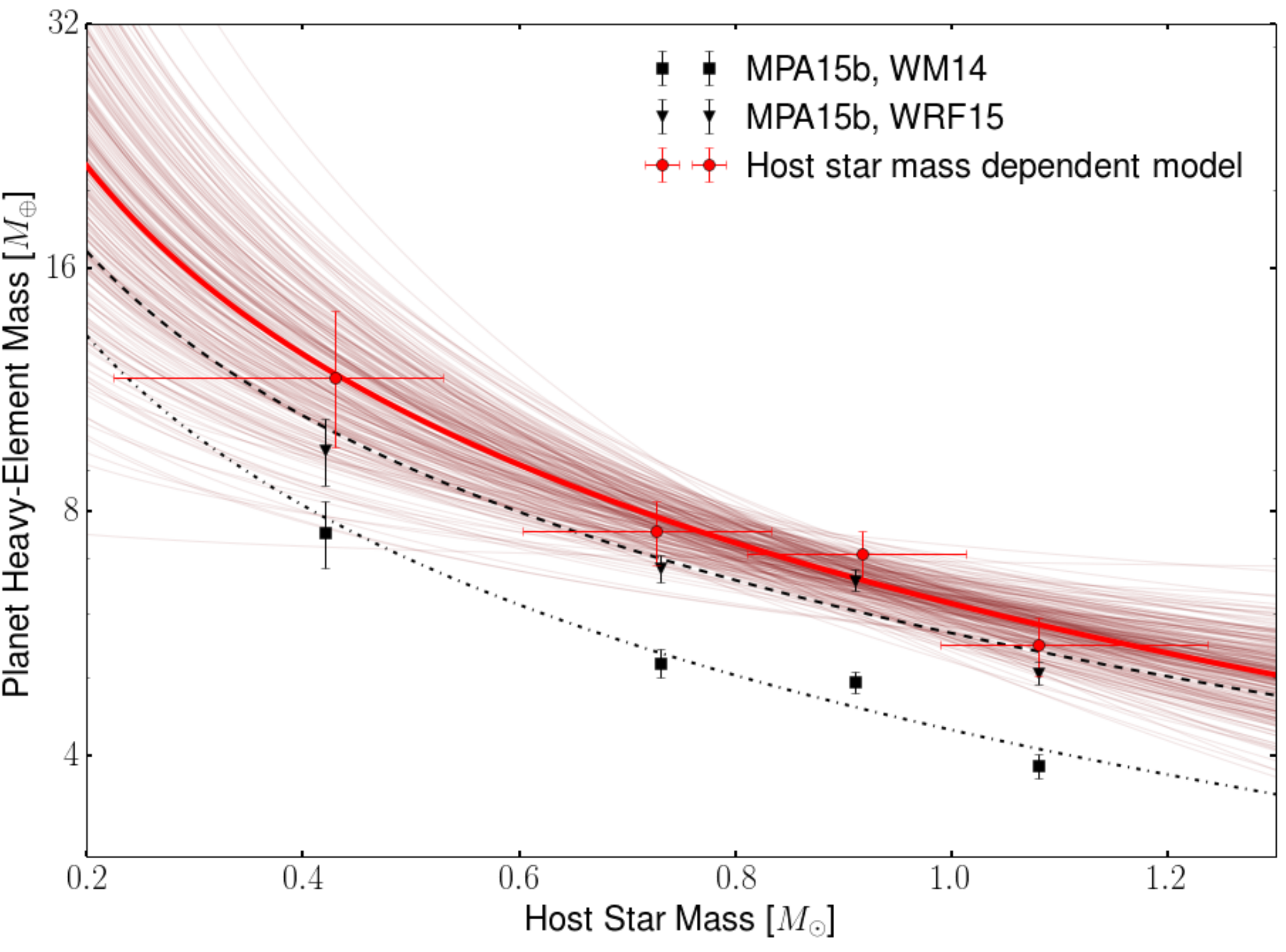}
    \caption{The total planetary heavy-element mass per star vs host star mass. Our current work is in red; black points are obtained from the best-fit WRF15 M-R relation (square) and from the best-fit WM14 relation (triangle). The points show the median heavy-element mass in that bin, with the error bars covering the central 68\%; the lines are fitted power laws to these four points. We fully incorporate M-R relation uncertainties as well as Poisson errors in obtaining our result: faded lines show different power law fits for each sampling of the M-R relation posteriors. Compared to MPA15b, we find a shallower slope but the overall trend of increasing planet heavy-element mass towards lower mass stars remains.}
    \label{Planet Heavy-Element mass}
\end{figure}

Our results are shown in Figure \ref{Planet Heavy-Element mass}. For these results, $N_m = 25$ and $N_b = 500$. The data shown is the median over $N_b$ and $N_m$ samples, with error bars representing the $15.9$ and $84.1$ percentiles. For each $T_{\text{eff}}$ bin, we plot the corresponding host star mass by taking the median of the masses of each star in the bin, with error bars again representing the $15.9$ and $84.1$ percentiles. We find that our results are consistent with the findings in MPA15b. The average planetary heavy-element mass increases from 5.4$\substack{+1.1 \\ -0.9} M_{\odot}$ around F stars to 11.7$\substack{+4.9 \\ -3.6} M_{\odot}$ around M stars. With respect to MPA15b, our error bars are roughly twice as large and our average masses slightly higher, but the overall trend is consistent.

\begin{table}
\centering
\begin{tabular}{c c c c c} 
 \hline
 Assumptions & M & K & G & F \\
 \hline\hline
 1) MPA15b & 11.2 $\substack{+4.9 \\ -3.6}$ & 7.4 $\substack{+1.6 \\ -1.3}$ & 7.0 $\substack{+1.2 \\ -1.0}$ & 5.4 $\substack{+1.1 \\ -0.9}$ \\ 
 2) $M_Z \propto \sqrt{M_p}$ & 10.6 $\substack{+4.2 \\ -3.1}$ & 6.7 $\substack{+1.5 \\ -1.3}$ & 6.2 $\substack{+1.2 \\ -1.0}$ & 4.9 $\substack{+1.1 \\ -0.9}$\\
 3) Snowline scaling + 2 & 8.4 $\substack{+3.1 \\ -2.3}$ & 5.9 $\substack{+1.3 \\ -1.1}$ & 6.1 $\substack{+1.1 \\ -1.0}$ & 5.0 $\substack{+1.1 \\ -0.9}$\\
 4) Incident flux dep. + 3 & 8.4 $\substack{+3.2 \\ -2.4}$ & 5.7 $\substack{+1.6 \\ -1.3}$ & 5.6 $\substack{+1.7 \\ -1.2}$ & 4.7 $\substack{+1.6 \\ -1.1}$ \\
 5) TTV + 3 & 5.9 $\substack{+3.2 \\ -1.9}$ & 3.9 $\substack{+1.1 \\ -0.9}$ & 3.9 $\substack{+0.9 \\ -0.8}$ & 3.1 $\substack{+0.9 \\ -0.7}$ \\ 
 6) TTV + 4 & 7.0 $\substack{+3.5 \\ -2.5}$ & 4.4 $\substack{+1.6 \\ -1.2}$ & 4.4 $\substack{+1.5 \\ -1.1}$ & 3.8 $\substack{+1.7 \\ -1.2}$ \\ 
 7) RV+TTV + 3  & 7.6 $\substack{+3.1 \\ -2.2}$ & 5.2 $\substack{+1.4 \\ -1.1}$ & 5.2 $\substack{+1.5 \\ -1.1}$ & 4.4 $\substack{+1.3 \\ -1.1}$ \\
 8) RV+TTV + 4 & 7.6 $\substack{+3.0 \\ -2.2}$ & 5.4 $\substack{+1.2 \\ -1.0}$ & 5.5 $\substack{+1.1 \\ -0.9}$ & 4.6 $\substack{+1.0 \\ -0.9}$ \\
 \hline
\end{tabular}
\caption{Average planet heavy-element mass per star, under different assumptions. Assumptions are described more explicitly in the text.}
\label{table:heavy element mass}
\end{table}

With the assumed model of the host star mass dependent M-R relation, our results are consistent within error with neither a scaling similar to the scaling of protoplanetary disk masses with host star mass, nor a flat trend. We fit a power-law to the results and find a power-law index of -0.9. We also fit a power-law to each of the $N_b$ individual samples of the planet population and M-R relation parameters. These are shown as faded lines in Figure \ref{Planet Heavy-Element mass} and give a rough visualization of the spread of this trend with host star mass. 

We have tested several combinations of parameters to see what would be consistent with a flat trend of planetary heavy-element mass with host star mass. Keeping $\left\{C_0, \gamma_0, \sigma_0\right\}$ fixed to their median values as listed in Table \ref{table:parameter posteriors}, we find that the following combination of $\left\{C_s = 1, \gamma_s = 1.6, \sigma_s = -5\right\}$ is sufficient to produce a flat trend. This results in a flat M-R relation for planets with $R > 1.3R_{\oplus}$ orbiting M dwarf ($M {\sim} 0.42 M_{\odot}$) stars. As seen in our model posteriors in Figure \ref{Cornerplot}, this set of parameters is well outside the $2 \sigma$ contours and is heavily disfavored by the current dataset.

\begin{figure}
    \centering
    \includegraphics[width=\columnwidth]{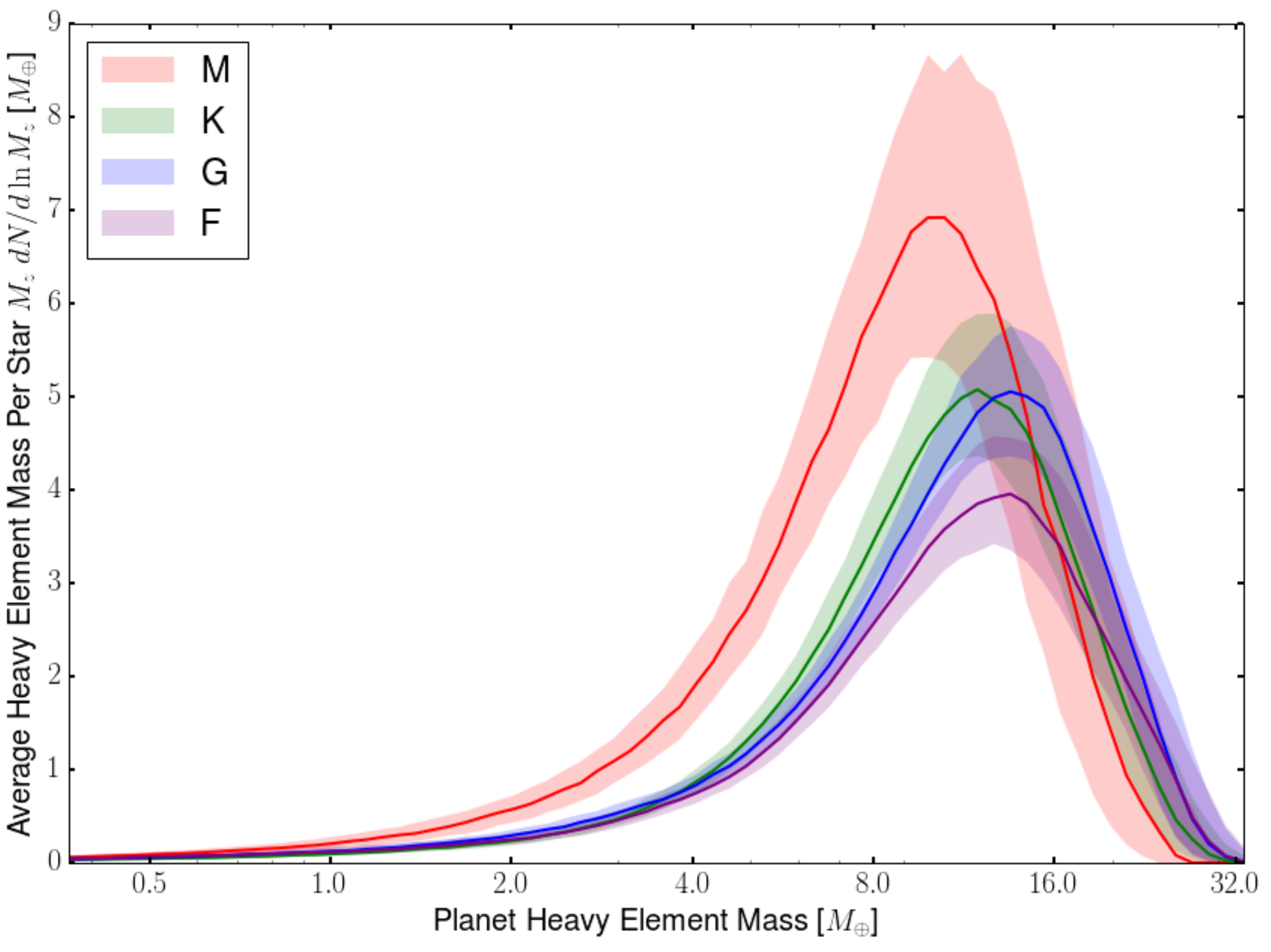}
    \caption{The average heavy element mass per star, binned and scaled by heavy element mass. The red curve is for planets around M dwarf stars, with the green, blue and purple curves representing K, G and F stars respectively. The shaded region shows the $68 \%$ region, obtained through bootstrap resampling and sampling of the M-R relation parameters. These results are for the third line in Table \ref{table:heavy element mass}, where we scale the heavy element mass as $\sqrt{M_p}$ and scale the period bounds with the location of the snowline. The integral under each curve gives the total heavy-element mass reported in Figure \ref{Planet Heavy-Element mass}. We find that the increased planetary heavy element mass around M dwarfs at short periods is due to higher occurrence of planets with $M_Z < 20 M_{\oplus}$.}
    \label{Unbinned Heavy Element Mass}
\end{figure}

In order to test the robustness of this trend in heavy-element mass, we would like to know to what extent the increase at lower host star mass is dependent on our assumptions. First, we test the assumption that heavy-element mass is equivalent to planet mass up to a limiting mass, chosen to be $22 M_{\oplus}$ in MPA15b due to the median core mass of giant planets. Recent evidence shows that the correlation between heavy-element mass and total mass for giant planets roughly scales as $M_Z \propto \sqrt{M}$, based on thermal and structural evolution models \citep{Thorngren2016ApJ}. This correlation is fit to masses above ${\sim} 20 M_{\oplus}$. We adopt this scaling for planets with $R > 4 R_{\oplus}$, which roughly corresponds to a mass of $20 M_{\oplus}$ in our model, with $M_Z \propto M$ below $4 R_{\oplus}$. To smoothly transition between these two functions, we adopt a logistic function with a transition point at $4 R_{\oplus}$ and a scale parameter of $1$. For giant planets with $R > 10 R_{\oplus}$, the M-R relation changes to become approximately flat \citep{ChenKipping2017ApJ}, and so we assume the heavy-element mass is flat past this radius. 

The second assumption we test is the period cut we apply on the planet sample. A planet with a period of 150 days has a significantly different irradiation depending on what type of star it is orbiting. By applying a cut of $2-150$ days regardless of host star mass, we are applying an uneven cut in planet irradiation. To account for this, we use the location of the snowline as a proxy for planet irradiation and use the scaling found in \citet{Ida&Lin2005ApJ}, where $a_{\text{snowline}} = 2.7 \text{AU} (M_*/M_{\odot})$. This is equivalent to a linear scaling of the period with host star mass, by Kepler's third law. We take $2-150$ days to be the period bounds for a host star with $1 M_{\odot}$, and scale both the inner and outer bound by $M_*/M_{\odot}$. For example, for a planet with a host star mass of $0.5 M_{\odot}$, it is only included in the sample if its period falls between $1-75$ days. 

For these two cases (where we also include the $M_Z$ scaling in the second case), along with our initial assumptions, we calculate the average heavy element mass per star in bins of heavy element mass, for each of our four host star mass bins. We then integrate this curve along the heavy element mass axis to obtain the total average heavy element mass per star. The results are shown in Table \ref{table:heavy element mass}. We find that scaling the heavy element mass as $\sqrt{M}$ decreases the total heavy element mass across all four host star mass bins. Scaling the period bounds with the location of the snowline decreases the heavy element mass for M and K stars, making the slope shallower. Despite changing our assumptions for the heavy element mass and period scaling, the trend of increasing heavy element mass towards lower host star mass is still present. Figure \ref{Unbinned Heavy Element Mass} shows the contribution of different heavy element mass regimes to the total heavy element mass per star; the integral of each curve in Figure \ref{Unbinned Heavy Element Mass} gives the total heavy element mass shown in Figure \ref{Planet Heavy-Element mass}.  Both the snowline and heavy element mass scalings are included in Figure \ref{Unbinned Heavy Element Mass}. We can see that the increase in the average heavy element mass for M dwarfs is due to planets with heavy element masses up to ${\sim}16 M_{\oplus}$ contributing more mass than those around F, G and K stars. Only at masses higher than $16 M_{\oplus}$ do planets contribute more heavy element mass for F, G and K stars compared to M stars.

\begin{figure}
    \centering
    \includegraphics[width=\columnwidth]{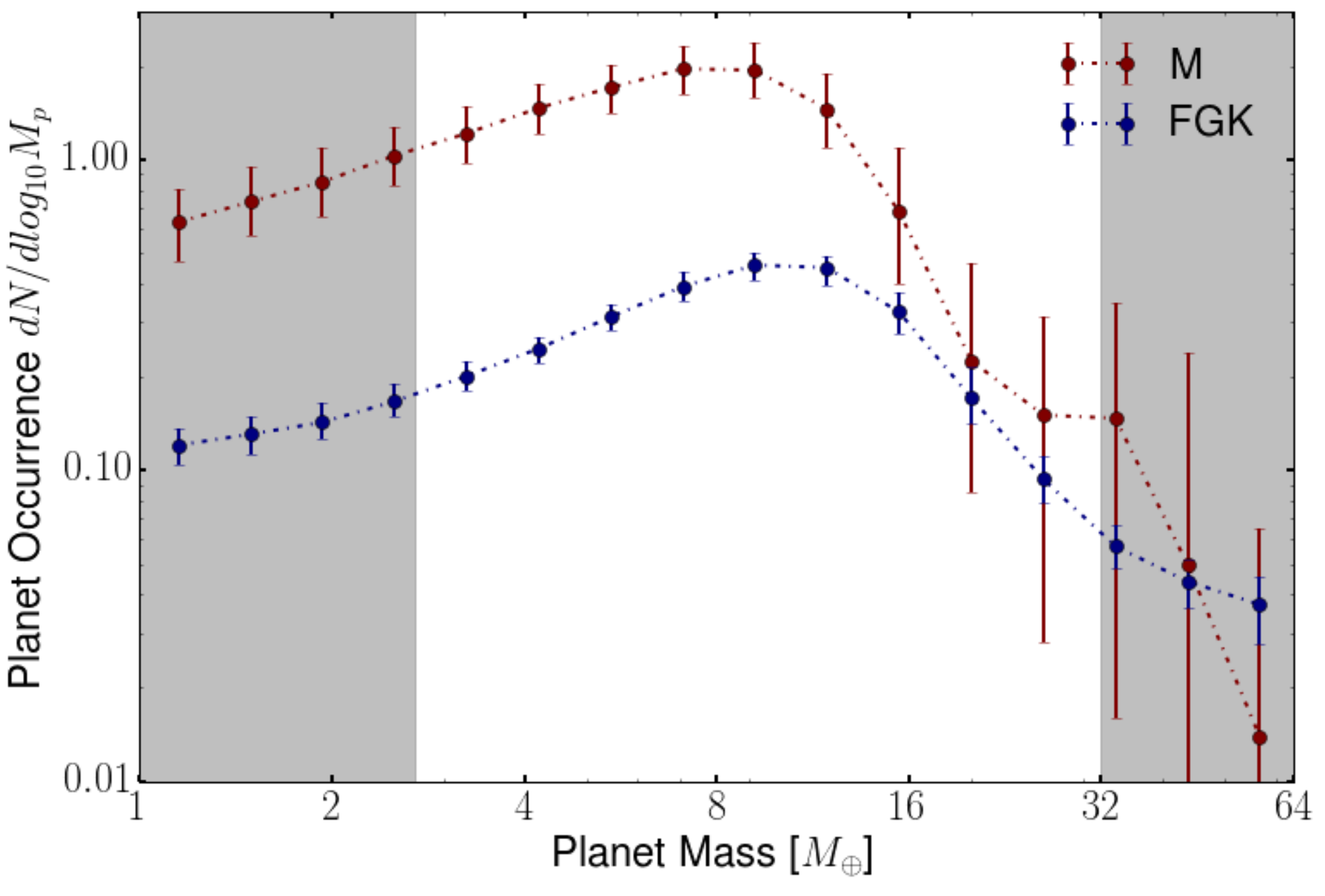}
    \caption{Mass distributions for \textit{Kepler} planets with short period orbits around M dwarfs (Red) and FGK dwarfs (Blue). Distributions were obtained by sampling from the probabilistic host star mass dependent M-R relation posteriors. Error bars correspond to the central 68\% of occurrence rates drawn at a given mass bin. The grey shaded regions indicate incompleteness of the sample: below 3 $M_{\oplus}$, the \textit{Kepler} sample is incomplete, and above 32 $M_{\oplus}$, the M-R relation we fit has to be extrapolated. The mass distribution for planets around M dwarfs peaks at a lower mass and has higher occurrence for less massive planets than the distribution for planets around FGK dwarfs.}
    \label{Mass distribution}
\end{figure}

Using similar methodology to calculating the planet heavy-element mass, we also calculate mass distributions for \textit{Kepler} planets around M stars as well as FGK stars, shown in Figure \ref{Mass distribution}. For this calculation, we use the period range scaling with the snowline, and we bin by planet mass as well as stellar $T_{\text{eff}}$. We find that planet occurrence is higher for planets with $3 < M/M_{\oplus} < 32$ around M dwarfs than FGK dwarfs. Beyond $32 M_{\oplus}$, we are limited by the small number of planets discovered around M dwarfs, as well as the limited radius range to which we fit the M-R relation. Below $3 M_{\oplus}$, the \textit{Kepler} sample is thought to be incomplete for the period range under consideration \citep{ChristiansenEt2015ApJ}.

We find an occurrence rate of 1.03$\substack{+0.24 \\ -0.20}$ planets per M dwarf with $M < 10 M_{\oplus}$ and $1 < P < 100$ d. Compared to the \textit{HARPS} M dwarf sample for the same mass and period bounds which found 0.88 planets per star \citep{BonfilsEt2013AAP}, our derived result is a factor of 1.17 higher but consistent within 1 sigma. We find 0.30$\substack{+0.19 \\ -0.12}$ planets per M dwarf with $10 < M/M_{\oplus} < 100$ and $1 < P < 100$ d, which is a factor of 6 higher than the \textit{HARPS} result of 0.05. We attribute this higher frequency of massive planets to our sampling of the masses of each planet, which has the effect of flattening the distribution. Since our masses are more uncertain than directly measured radial velocity masses, planets with smaller radii at higher occurrence rates will have some significant posterior probability at higher masses.

\subsection{Minimum-mass extrasolar nebula}

The minimum-mass solar nebula (MMSN) is an estimate of how much mass must have been in the solar protoplanetary disk to form the planets had they formed in situ. \citet{Kuchner2004ApJ} first showed how one could extend the concept of the MMSN to exoplanets, using a sample of radial velocity planets in order to calculate the minimum-mass extrasolar nebula (MMEN). More recently, \citet{Chiang&Laughlin2013MNRAS} calculated the surface density profile of the MMEN using the \textit{Kepler} sample of transiting planets. They find a similar power-law slope compared to the MMSN (-1.6 and -1.5, respectively) and an overdensity compared to the MMSN of about a factor of five. However, their primary result has several limitations. The M-R relation they use is the one derived in \citet{LissauerEt2011Nature}: a power-law fitted to six solar system planets, which has since been greatly improved upon by utilizing the sample of transiting exoplanets with mass measurements. The host star mass for each individual Kepler planet is not taken into account, neither for the calculation for orbital radius nor for the M-R relation. Finally, the occurrence rate of each individual planet is not factored in. Here, we redo the MMEN calculation, making several improvements over the initial work.

We use the \textit{Kepler} occurrence rates, planet heavy-element mass and host star mass samples as described in section 3.3 for the following calculation. For each sample, the solid surface density of the planet is given by the following equation:

\begin{equation}
\Sigma_{\text{solid}} = \frac{M_{Z}}{2\pi a^2}
\end{equation}

\begin{figure}
    \centering
    \includegraphics[width=\columnwidth]{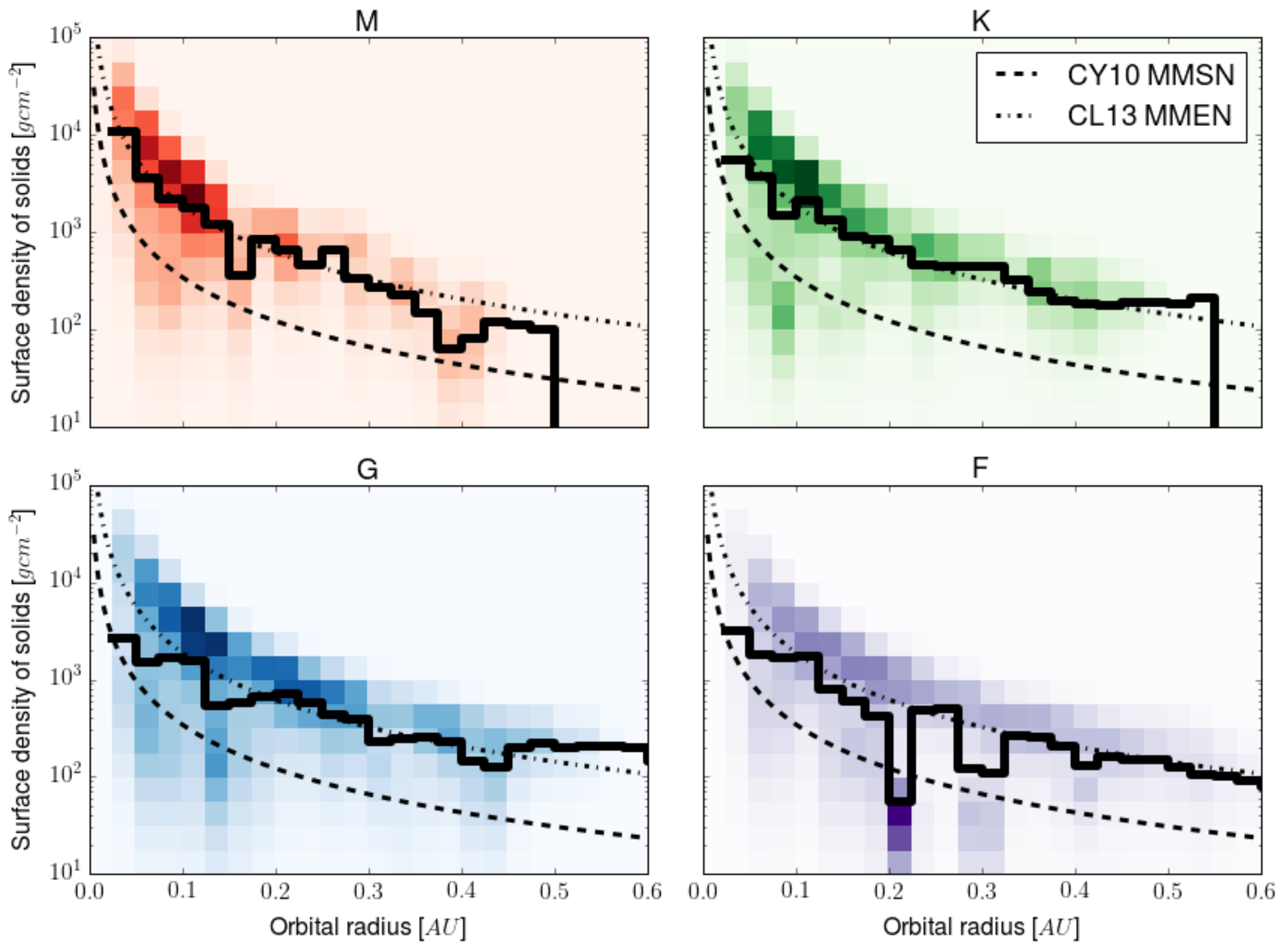}
    \caption{A 2D weighted histogram of the MMEN for different types of stars, using the \textit{Kepler} sample. Samples are drawn for each KOI using our host star mass dependent M-R relation posteriors, and weighted by the occurrence rate of the KOI. The solid black line indicates the cumulative center of each orbital radius bin, and the dashed and dot-dashed lines show the results from the \citet{Chiang&Youdin2010AREPS} MMSN and the \citet{Chiang&Laughlin2013MNRAS} MMEN. We find that the slope of the surface density profile of the MMEN is steeper for M dwarfs, resulting in more mass at short orbits around M dwarfs.}
    \label{MMEN}
\end{figure}

In Figure \ref{MMEN}, we plot a 2D weighted histogram of the MMEN surface density of solids and semimajor axis for M, K, G, and F dwarfs separately, where the weights are the occurrence rates of the KOIs. We find that the surface density of solids at short orbits is higher for M dwarfs than it is for FGK dwarfs. Between F, G and K dwarfs there is little difference, mirroring Figure \ref{Unbinned Heavy Element Mass}. Compared to \citet{Chiang&Laughlin2013MNRAS}, we find that the surface density profile of the MMEN for FGK dwarfs exhibits a shallower slope with a power-law index ranging from $-0.9$ to $-1.2$, whereas the power-law index for M dwarfs, $-1.8$, more closely matches their result of $-1.6$. We find a similar normalization for the surface density of solids, which is several factors higher than that of the MMSN. While the MMEN suggests a higher surface density of solids at short orbits for disks around M dwarfs, migration likely plays a pivotal role in this trend and thus the MMEN may not reflect the true initial protoplanetary disk surface density profile \citep{MuldersEt2015bApJ}. This is supported by recent work from \citet{Raymond&Cossou2014MNRAS}, which showed that the surface density profiles of multiple planet systems have a wide range of power-law slopes.

\section{Discussion}

\subsection{Model selection}

In order to test whether we are justified in adding host star mass dependence to the M-R relation, we would like to estimate the predictive accuracy of models with and without host star mass dependence. Cross-validation is a robust method of evaluating the predictive accuracy of a model, but requires multiple model fits and is computationally expensive. Luckily, several alternatives exist which approximate cross-validation and are simpler to compute. For the purposes of this paper, we choose Watanabe-Akaike information criterion (WAIC) to be our predictive measure of choice \citep{Watanabe2012CS}. Similar to other information criterion, WAIC relies on computing the log predictive density for the data sample used to fit the model, and applying a bias correction to estimate the log predictive density for a hypothetical new data sample. Unlike alternatives such as AIC and DIC, WAIC is fully Bayesian in the sense that it averages over the posterior distributions, rather than relying on point estimates \citep{Gelman2014STAT}. Given that the posteriors for our six parameters are not all normally distributed ($\sigma_s$ in particular), this makes WAIC an appealing choice.

\begin{table}
\centering
\begin{tabular}{c c c c} 
 \hline
 Model & RV & TTV & RV+TTV \\
 \hline\hline
 No $M_*$ dependence & -14.8 $\pm$ 4.6 & -8.4 $\pm$ 2.4 & -17.3 $\pm$ 4.6 \\
 $C_s$ & -14.8 $\pm$ 2.9 & -5.2 $\pm$ 2.1 & -17.0 $\pm$ 2.7 \\
 $\gamma_s$ & -11.7 $\pm$ 2.1 & -7.4 $\pm$ 2.2 & -11.9 $\pm$ 2.6 \\
 $\sigma_s$ & -7.5 $\pm$ 4.0 & -2.8 $\pm$ 1.6 & -9.4 $\pm$ 3.7 \\
 $C_f, \gamma_f, \sigma_f$ & -5.9 $\pm$ 3.8 & 2.5 $\pm$ 2.2 & -8.0 $\pm$ 3.3 \\
 $C_s, C_f, \gamma_s, \gamma_f, \sigma_s, \sigma_f$ & 4.3 $\pm$ 3.2 & 40.1 $\pm$ 6.3 & -1.7 $\pm$ 4.7 \\
 \hline
\end{tabular}
\caption{Difference in Watanabe-Akaike information criteria (WAIC) compared to the standard six-parameter host star mass dependent model, along with the error in the difference. Results are shown for three datasets and six different models. Negative numbers favor the model in question over the six-parameter model. We find that models with fewer parameters are favored.}
\label{table:waic}
\end{table}

We compute WAIC for seven different models: our six-parameter host star mass dependent M-R relation, a three-parameter model with no host star mass dependence, three four-parameter models that correspond to adding one of $C_s$, $\gamma_s$, and $\sigma_s$ to the three-parameter model, our six-parameter incident flux dependent model, and our nine-parameter model with both incident flux and host star mass dependence. We then calculate the difference in WAIC between the six-parameter model and each other model, along with the error in the difference. We repeat this for the TTV and RV+TTV datasets discussed in section 3.3. Our results are shown in Table \ref{table:waic}. Lower values for a given model indicate better predicted out-of-sample fit compared to the six-parameter model. We find that generally, models with fewer parameters have a higher predicted out-of-sample fit compared to our standard six-parameter model. Our results for the incident flux dependent model vary between datasets, with incident flux dependence being most strongly favored over host star mass dependence for the combined RV+TTV dataset, but slightly disfavored for the TTV only dataset. For the RV and TTV datasets, the nine-parameter model with both host star mass and incident flux dependence is disfavored, whereas for the combined RV+TTV dataset, the nine-parameter model is slightly favored but with a large error in the difference. Given that the error of these estimates is often comparable to the calculated differences, nothing definitive can be said about which model should be preferred. This strengthens our conclusion that there is no evidence in the current M-R dataset for host star mass dependence in the M-R relation.

\subsection{Limitations}

\subsubsection{Data set}

The limited number of transiting planets with RV mass measurements and their uneven occupation of parameter space is perhaps the most significant factor limiting this work. $48 \%$ of planets in our sample have a host star mass between $0.9-1.1 M_{\odot}$, and only six planets have a host star mass below $0.7 M_{\odot}$. Given our model parametrization, where host star dependence scales as $\text{ln}M_*$, the difference between the coefficient of the host star mass dependent parameters is ${\sim}0.2$ between $0.9-1.1 M_{\odot}$, but ${\sim}0.6$ between $0.5-0.9 M_{\odot}$. Under this scaling, the M-R relation is most significantly different for planets around M dwarf stars, but we only have six such planets. Furthermore, $58 \%$ of planets in our sample have radii between $1.5-4.0 R_{\oplus}$. More planets with radii between $4.0-8.0 R_{\oplus}$ would further constrain the slope of the power-law and allow a more substantive investigation into whether the scatter of the M-R relation changes with radius. These problems are twofold: first, there is the detection of transiting planets within our parameters of interest, and secondly, the radial velocity follow-up of these planets. Both issues need to be addressed, although mass-radius studies are more sensitive to the number of planets subjected to RV follow-up given the wealth of data provided by \textit{Kepler}.

Fortunately, there are several surveys and experiments scheduled to become operational in the near future that will alleviate these issues. The Transiting Exoplanet Survey Satellite (\textit{TESS}), scheduled to launch in 2018, is an all-sky NASA-sponsored mission designed to monitor ${\sim} 200,000$ of the brightest nearby stars. \textit{TESS} is expected to find ${\sim} 400$ planets with $R < 2 R_{\oplus}$ hosted by M dwarfs \citep{Sullivan2015ApJ}, compared to ${\sim} 130$ found by \textit{Kepler}. On the RV follow-up side, several high-resolution spectrographs (e.g. \textit{MAROON-X}, \textit{SPIRou}, \textit{ESPRESSO}) are under development and planned to coincide with \textit{TESS}. There are also efforts currently underway to search for planets around nearby M dwarfs (e.g. \textit{CARMENES}, \textit{MEarth}, \textit{IR RV} \text{spectrograph}). Since \textit{TESS} is scheduled to observe the brightest nearby M dwarfs, RV follow-up of planets around M dwarfs in the era of TESS will be more feasible. As shown in section 3.3, a future dataset of planets from \textit{TESS} with radial velocity mass measurements may be able to distinguish between M-R relations with and without host star mass dependence.

\subsubsection{Physical Basis}

Given that we are empirically fitting the M-R relation, there is no shortage of parametrizations we could have considered. Much like the decision to characterize the M-R relation as a power law, the decision to scale the M-R relation parameters by the natural log of host star mass was based on simplicity and intuitive understanding, rather than any physical basis. There is no reason to think that this scaling should be physically preferred over a power law scaling, for instance. Our motivation for this paper was to allow for host star mass dependence and see how much information is in the current dataset. For this reason, we do not consider alternate parametrizations.

Ultimately, M-R relations should move away from strictly empirical relations and toward a physically motivated distribution. One possible step in this direction is to use a mixture model. Results from the \textit{Kepler} survey show a gap in the radius distribution of small planets at short orbital periods \citep{FultonEt2017ApJ}. This bimodal distribution is thought to arise from two separate planet populations: those with significant H/He envelopes and those without. The gap is consistent with evidence that planets below ${\sim} 1.6 M_{\oplus}$ having densities consistent with a purely rocky composition \citep{Rogers2015ApJ, Weiss&Marcy2014ApJL, DressingEt2015ApJ}. Modeling these two planet populations separately using a mixture model would give each planet a probability of falling into either population, depending on its radius. This would be an improvement over current efforts that model a break in the power-law around $1.6 M_{\oplus}$, as it would account for the overlap between these two populations.

\subsubsection{Mass Conditioned on Radius}

In this paper we parameterize the M-R relation in terms of mass as a function of radius, in order to directly apply the relation to the \textit{Kepler} sample of planets, which generally lack mass measurements. Framing the M-R relation as \textit{M(R)} allows mass estimates of \textit{Kepler} planets to be readily obtained. Furthermore, the radius measurements for the planets in our sample are generally much more precise than the mass measurements. However, one could argue that mass is the more fundamental physical quantity and the relation should be cast as radius as a function of mass. A \textit{R(M)} relation would also be applicable to the sample of microlensing planets, which have mass constraints but no radius measurements. If desired, one could easily obtain radius as a function of mass by switching mass and radius in Equation (\ref{eq:1}) and fitting for a new set of parameters. Ultimately, a combined joint mass-radius distribution would solve this issue, which would allow one to obtain either relation.

\subsubsection{Selection Effects}

In using the sample of transiting planets with radial velocity mass measurements, we are subject to a host of poorly characterized selection effects. For example, the choice to follow up a \textit{Kepler} planet is not completely transparent, and while guidelines exist, it is often ultimately a human decision. Planets orbiting M dwarfs are typically not favored to be chosen for RV follow-up, due to their intrinsic faintness and the desire to characterize planets around solar analogs. Furthermore, upper limits when a planet is not detected are not always published, which could lead to a bias towards more massive planets as they are more likely to be detected. The heterogeneity of the data set also poses a problem: some planets in the sample were discovered by both radial velocity and transit methods independently, and we do not restrict our catalog to one specific survey or program such as \textit{Kepler}. Finding a way to model these selection effects is a difficult task, but necessary to remove any potential biases.

\subsubsection{Incident Flux Dependence}

In section 3.2 we explored adding incident flux dependence to the M-R relation. We found that with the current dataset, we are unable to distinguish between the need for host star mass dependence and the need for incident flux dependence. This is in agreement with previous work by \citet{WeissEt2013ApJ} which found weak dependence of planet radius on incident flux for small planets.  Despite this, incident flux is a key parameter to take into account when constraining planet composition distributions. The incident flux on a planet can affect whether or not a planet retains its atmosphere and the thermal evolution of a planet, particularly for those in close-in orbits \citep{ScaloEt2007AsBio}. An increased sample of planets with mass and radius measurements will warrant revisiting this incident flux dependence. Additionally, the sample of transiting planets with TTV mass measurements will provide another avenue to explore this dependence, given that TTV techniques prefer planets on longer periods than RV methods \citep{Mills2017ApJL}.

\section{Conclusion}

We have modeled host star mass dependence in the planet M-R relation by introducing three new parameters to the probabilistic M-R relation first established in \citet{WolfgangEt2016ApJ}. We fit the model to the current sample of transiting planets with RV measured masses and find that the host star mass dependent parameters are consistent with zero and there is no strong evidence for host star mass dependence in the M-R relation. We have tested the observed trend in \citet{MuldersEt2015bApJ} of increasing planetary heavy element mass towards lower mass stars and have found this trend to be robust against many of their assumptions. This trend also manifests itself in the minimum-mass extrasolar nebula, with the surface density profile for M dwarfs exhibiting a steeper slope than that of FGK dwarfs. This work provides a framework for including host star mass dependence in the M-R relation, and can be revisited when more planets around M dwarfs have their masses and radii characterized by upcoming surveys such as \textit{TESS} and subsequent radial velocity follow-up.

\bigskip

We thank our anonymous referee for providing valuable feedback and suggesting additions that rounded out the paper. We also thank Angie Wolfgang, Daniel Fabrycky, Emily Gilbert and Gregory Gilbert for useful discussion. LAR gratefully acknowledges support from NASA Exoplanet Research Program grant NNX15AE21G and from NSF FY2016 AAG Solicitation 12-589 award number 1615089.

\bibliography{exoplanets}

\clearpage

\end{document}